\documentclass[twocolumn,showpacs,preprintnumbers,amsmath,amssymb,floatfix]{revtex4}
\usepackage{graphicx}
\usepackage{slashed}
\usepackage{color}
\begin{document}

\title{Coupled $K^+\Lambda$ and $K^0\Lambda$ photoproduction off the nucleon:
   Consequences from the recent CLAS and MAMI data and the $N(1680)P_{11}$ narrow state}

\author{T. Mart}
\email[]{terry.mart@sci.ui.ac.id}

\affiliation{Departemen Fisika, FMIPA, Universitas Indonesia, Depok 16424, Indonesia}


\begin{abstract}
The new $\gamma n\to K^0\Lambda$ data obtained from the CLAS and MAMI 
collaborations are analyzed by employing an effective Lagrangian 
method. The constructed model can describe all available experimental 
data in both $\gamma p \to K^+\Lambda$ and $\gamma n\to K^0\Lambda$ 
channels, simultaneously. The background part of the model is built
from the appropriate intermediate states involving the nucleon, kaon, and
hyperon exchanges, whereas the resonance part is constructed from the 
consistent interaction Lagrangians and propagators. To check the performance 
of the model a detailed comparison between the calculated observables and 
experimental data in both isospin channels is presented, from which a nice 
agreement can be observed. The discrepancy between the CLAS and MAMI data
in the $\gamma n\to K^0\Lambda$ channel is analyzed by utilizing three
different models; M1, M2, and M3 that fit the CLAS, MAMI, and both CLAS and 
MAMI data sets, respectively.
The effect of this discrepancy is studied by investigating the significance
of individual nucleon resonances and the predicted beam-target helicity asymmetry 
$E$ that has been measured by the CLAS collaboration recently. It is found that the 
$N(1720)P_{13}$, $N(1900)P_{13}$, and $N(2060)D_{15}$ resonances are significant  
for improving the agreement between model calculation and data. This result is
relatively stable to the choice of the model. The helicity asymmetry $E$ can be 
better explained by the models M1 and M3. Finally, the effect of 
the $N(1680)P_{11}$ narrow resonance on the cross section of both isospin channels
is explored. It is found that the effect is more sensitive in the $\gamma n\to K^0\Lambda$ 
channel. In this case the model M3, that fits both CLAS and MAMI data, yields a 
more realistic effect.
\end{abstract}

\pacs{13.60.Le, 14.20.Gk, 25.20.Lj}

\maketitle
\section{INTRODUCTION}
Among the six possible isospin channels in kaon photoproduction, the
$\gamma p \to K^+\Lambda$ channel is the first channel and most studied 
channel during the last six decades. This is understandable because from both 
theoretical and experimental points of view this channel is the simplest, 
although accurate data could only be obtained in the last two decades
after the use of continuous and heavy duty accelerators along with 
the unprecedented precise detectors. As a consequence, the $K^+\Lambda$ 
channel has the largest experimental database collected from different
measurements performed in the modern accelerator facilities, such as 
CEBAF in Jefferson Lab, SPring8 in Osaka, MAMI in Mainz, ELSA in Bonn 
and ESRF in Grenoble. These accurate data allow for detailed analyses
of certain nucleon and delta resonances in kaon photoproduction that are not
available in other meson photoproductions due to their large kaon
branching ratios.
Further information to this end can be found, e.g., 
in our recent reports \cite{Mart:2017mwj,Clymton:2017nvp}.

The history of kaon photoproduction presumably started with the theoretical  
analysis by Kawaguchi and Moravcsik more than 60 years ago \cite{masasaki}, after 
the work of Chew, Goldberger, Low, and Nambu (known later as the CGLN amplitude) 
\cite{chew}. In their analysis, Kawaguchi and Moravcsik considered all six 
possible isospin channels by using only three Feynman diagrams of the Born terms, 
in spite of the fact that no experimental data were available at that time. Furthermore, 
the parities of the $\Lambda$ and $\Sigma^0$ hyperons were also not known, 
although they decisively control the cross section magnitude. 
Therefore, it is safe to consider the pioneering work of Thom in 1966 \cite{thom} 
as a more serious attempt to this end, since both Born and resonance terms
had been considered, while coupling constants in the Born term and some resonance
parameters in the resonance term were fitted to 58 experimental data points (46 points 
of differential cross sections and 12 points of $\Lambda$ polarization).

The interest in the kaon photoproduction temporarily declined after the midseventies,
particularly due to the lack of new experimental data. However, with the construction of 
new generation of high duty-factor accelerators that provide continuous, high 
current, and polarized photon beams in the energy regime of a few GeV in Mainz (MAMI), 
Bonn (ELSA), Newport News (CEBAF), Grenoble (ESRF), and Osaka (SPring8), along with 
the unprecedented precise detectors, the interest in this topic revived. Indeed, we notice 
that nearly a decade before the operation of these facilities a number of significant 
theoretical analyses \cite{abw,williams,Adelseck:1990ch} had been made to predict the 
most relevant observables measured in these facilities.

In 1994 a new set of data on $K^+\Lambda$ photoproduction appeared from the SAPHIR collaboration 
\cite{bdt}. These data sparked new analyses of kaon photoproduction ranging from quark models to 
the chiral perturbation theory \cite{zpli,rubin,Cheoun:1996kn,Steininger:1996xw}. However, more 
precise cross section data obtained by the same collaboration in 1998 \cite{Tran:1998} drew 
more attention because they show a clear structure near $W=1.9$ GeV. This structure was
interpreted in Ref.~\cite{kaon-maid} as an evidence of a missing $D_{13}(1895)$ nucleon 
resonance, although different interpretations were also possible 
\cite{Janssen:2001pe,MartinezTorres:2009cw}. The structure was later realized as the effect
of the $P_{13}(1900)$ nucleon resonance contribution, instead of the $D_{13}(1895)$ one \cite{Mart:2012fa}.
This conclusion corroborated the previous finding of the Bonn-Gatchina group \cite{Nikonov:2007br}.

The SAPHIR collaboration finally 
improved their data in 2003 \cite{Glander:2003jw}, where almost 800 data points were 
extracted from their experiment. In spite of the noticeable improvement, for $W\gtrsim 1.7$ GeV
the SAPHIR cross section data exhibit substantial discrepancy with the CLAS ones published two
years later \cite{Bradford:2005pt}. The problem of data discrepancy was intensively discussed
in the literature. It is important to note that the KAON-MAID model \cite{kaon-maid,Mart:2000jv}
was fitted to the SAPHIR 1998 data \cite{Tran:1998}.
The use of the SAPHIR and CLAS data, individually or simultaneously,
leads to different extracted resonance parameters and eventually different conclusions 
on the ``missing resonances'' in the process \cite{Mart:2006dk}. 
The difference between the two data sets has been studied by using an
energy-independent normalization factor in each of the data sets 
\cite{Sarantsev:2005tg}. It was found that in both cases the inclusion of this 
factor results in an increase of $\chi^2$ and, as a consequence, it was concluded that the 
consistency problem did not originate from an error in the photon flux normalization. 
However, subsequent measurements of the $K^+\Lambda$ photoproduction cross section 
\cite{Sumihama06,Hicks_2007,mcCracken} are found to be 
more consistent with the result of CLAS 2006 collaboration  \cite{Bradford:2005pt}.
After that, most analyses of the $K^+\Lambda$ photoproduction use the SAPHIR data only 
up to $W=1.7$ GeV \cite{Mart:2017mwj,Clymton:2017nvp,Mart:2012fa,Maxwell:2007zza,Borasoy:2007ku}.

So far, kaon photoproduction has been proven as an indispensable tool for 
investigating nucleon resonances, especially for those that have larger branching
ratios to the strangeness channels. The selection of the contributing nucleon resonances 
in the process is commonly left to the data through the $\chi^2$ minimization. 
To improve this selection process  a Bayesian inference method  
has been proposed to determine the most important
nucleon resonances in this process in a statistically solid way \cite{DeCruz:2011xi}.
Very recently, the least absolute shrinkage and selection operator method
combined with the Bayesian information criterion had been used to determine
the most important hyperon resonances in the ${\bar K}N\to K\Xi$
reaction  \cite{Landay:2019}. It was found that ten resonances, out of the 21 hyperon 
resonances with spin 7/2 listed by the Particle Data Group (PDG) \cite{pdg},
may potentially contribute to this reaction.

Nevertheless, a more 
conclusive result can be obtained from the coupled-channels analysis,
since in the latter all possible decays of the nucleon resonance are
considered by calculating all relevant scattering and photoproduction 
processes simultaneously, whereas the unitarity is preserved automatically.
We note that there have been a number of coupled-channels analyses performed in
the last decades to simultaneously describe the $\pi N$, $\pi\pi N$, $\eta N$, $K\Lambda$
scattering and photoproduction \cite{Feuster:1998cj,Shklyar:2014kra,Chiang:2001pw,%
Julia-Diaz:2006is,Anisovich:2011fc,Workman:2012jf,kamano,%
Manley:2003fi,Tiator:2018pjq,Hunt:2018mrt,Ronchen:2018ury}.
Most of them became the important source of information tabulated in the 
Review of Particle Properties of the PDG \cite{pdg}. The Giessen coupled-channels 
model is based on the covariant Feynman diagrammatic approach and makes
use the $K$-matrix formalism \cite{Feuster:1998cj,Shklyar:2014kra}. The same method is 
also used by the EBAC-JLab group \cite{Julia-Diaz:2006is,kamano}. The 
$K$-matrix approach is also adopted by the Bonn-Gatchina \cite{Anisovich:2011fc} 
and the GWU/SAID \cite{Workman:2012jf} groups,
albeit with the Breit-Wigner parametrization in the resonance part.
The KSU group \cite{Manley:2003fi,Hunt:2018mrt} used the generalized 
energy-dependent Breit-Wigner parametrization and considered the nonresonance
backgrounds consistently. The J\"ulich-Bonn group extended their dynamical coupled-channels
model to the $K^+\Lambda$ photoproduction and fitted in total nearly 40000 data points
\cite{Ronchen:2018ury}.
We also note that in the MAID model the unitarity is
fulfilled by introducing the unitary phase in the Breit-Wigner amplitude in order
to adjust the total phase such that the Fermi-Watson theorem can be satisfied
\cite{Tiator:2018pjq}. 

In contrast to the $\gamma p \to K^+\Lambda$ channel, 
the $\gamma n\to K^0\Lambda$ process is not easy to measure since
the latter uses a neutron as the target. Because the neutron
is unstable, clever techniques are required to replace it with a 
certain nucleus that behaves as a neutronlike target and to suppress the
contribution from the rest of the nucleons inside the nucleus. Thanks
to the recent advancements in target, detector and computational 
technologies, most of the problems to this end can be overcome and 
precise data in the $K^0\Lambda$ channel have just been available
\cite{Compton:2017xkt,Akondi:2018shh}.

Besides being useful in the analysis of other baryon resonances, e.g.,
missing \cite{kaon-maid}  and narrow \cite{mart-narrow} resonances,
phenomenological models of the $\gamma p \to K^+\Lambda$ channel
are also needed for use in the investigation of hadronic coupling
constants \cite{Mart:2013ida}, hypernuclear production \cite{hypernuclear}, 
and the Gerasimov-Drell-Hearn sum rule \cite{gdh-kaonmaid}. Extending 
the model to the finite $Q^2$ region (using the virtual photon instead 
of the real one) enables us to explore the charge distribution of kaons 
and hyperons via kaon electroproduction \cite{Mart:1997cc}, which is not 
possible in the case of pion or eta electroproduction.

On the other hand, photoproduction of neutral kaon $\gamma n\to K^0\Lambda$
also plays an important role in hadronic physics. Theoretically, this channel
can be related to the $\gamma p \to K^+\Lambda$ one by utilizing the isospin
relations in the hadronic coupling constants \cite{Mart:1995wu}. 
Thus, the $K^0\Lambda$ photoproduction can serve 
as a direct check of isospin symmetry. By using a multipole formalism for the
resonance terms this neutral kaon photoproduction can also be used to extract
the neutron helicity amplitudes $A_{1/2}(n)$ and $A_{3/2}(n)$. We note that
these amplitudes are also listed in the Review of Particle Properties of
PDG \cite{pdg}. Extending the photoproduction model 
to the electroproduction one allows us to assess the electromagnetic form 
factors of the neutral kaons and hyperons, which have been predicted by 
a number of theoretical calculations \cite{k0formfacttor}.

This paper provides a report on the simultaneous analysis of  the 
$\gamma p \to K^+\Lambda$ and $\gamma n\to K^0\Lambda$ channels 
by using a covariant isobar model. The model is based on our previous
covariant model constructed by using an effective Lagrangian 
method and fitted to nearly 7400 $K^+\Lambda$ data points \cite{Clymton:2017nvp}. 
The model has been updated to include the latest double polarization data
from the CLAS 2016 collaboration \cite{paterson}. Finally, we extend the
model to include the new $\gamma n\to K^0\Lambda$ data. This is performed by
using the isospin symmetry relation for the hadronic coupling constants
in the hadronic vertices and a number of parameters obtained from PDG estimates 
\cite{pdg} in the electromagnetic vertices. There are 18 nucleon resonances
included in this model. Their proton transition magnetic moments $g_{\gamma pN^{*+}}$ 
can be obtained by fitting the predicted observables to the $\gamma p\to K^+\Lambda$ 
data. Analogously, the neutron magnetic moments $g_{\gamma nN^{*0}}$ can be extracted 
by using the $\gamma n\to K^0\Lambda$ data. 

We have noticed that a recent study \cite{Kim:2018qfu} on the neutral kaon 
photoproduction, $\gamma n\to K^0\Lambda$, has been performed within a 
similar framework to the present study. The effective Lagrangian method was
utilized to construct the resonance amplitude, whereas the Regge formalism
was used to describe the background term. The hadronic coupling constants were 
obtained from the quark model prediction. Comparison of the
predicted cross sections with experimental data exhibits a good agreement.
However, only the $\gamma n\to K^0\Lambda$ channel is considered in this work
\cite{Kim:2018qfu}. Furthermore, at $W\approx 1.75$ GeV contribution of
the nucleon resonances seems to be too strong, especially in the forward
angle direction. The new MAMI 2018 data were not included, because they
were not available before the time of publication of Ref.~\cite{Kim:2018qfu}.
Our present analysis provides important improvement to this model, i.e.,
including the $\gamma p \to K^+\Lambda$ channel, simultaneously, using the
effective Lagrangian method for both background and resonance terms in 
a consistent fashion, and incorporating more experimental data. Furthermore,
in this work we also analyze the new CLAS data on beam-target helicity asymmetry 
$E$ in the $\gamma n\to K^0\Lambda$ process, the significance of individual nucleon
resonances in both isospin channels, and investigate the influence of 
the $P_{11}$ narrow resonance on both channels. 

This paper is organized as follows. In Sec.~\ref{sec:formalism} we present the 
formalism used in our model. This includes the interaction Lagrangian, 
propagators, the isospin symmetry that relates the $\gamma p \to K^+\Lambda$ 
to the $\gamma n\to K^0\Lambda$ channels, the nucleon and hyperon resonances used 
in the model, as well as the observables used in the fitting database. Furthermore,
we also briefly review the technique to calculate the observables and some issues regarding 
hadronic form factors. In Sec.~\ref{sec:result} we present 
and discuss the result of our calculation and compare it with
the available experimental data. Extensive discussion on the helicity asymmetry 
$E$, the significance of individual nucleon resonances, and the influence of 
the $P_{11}$ narrow resonance on both channels is also presented in this 
section. Finally, we summarize our work and 
conclude our findings in Sec.~\ref{sec:summary}.

\section{FORMALISM, RESONANCES, AND DATA}
\label{sec:formalism}

\subsection{The interaction Lagrangian}
The interaction Lagrangians used in the present work can be found
in our previous analyses \cite{Mart:2013ida,Mart:2015jof,Clymton:2017nvp}.
For the Born terms the Lagrangians have been also given in 
many literatures, e.g., in Ref.~\cite{Kim:2018qfu}.
Nevertheless, for the convenience of the readers and to avoid confusion,
because different notations are commonly used, 
in the following we briefly review the formulas.

\begin{figure*}
  \includegraphics[scale=0.60]{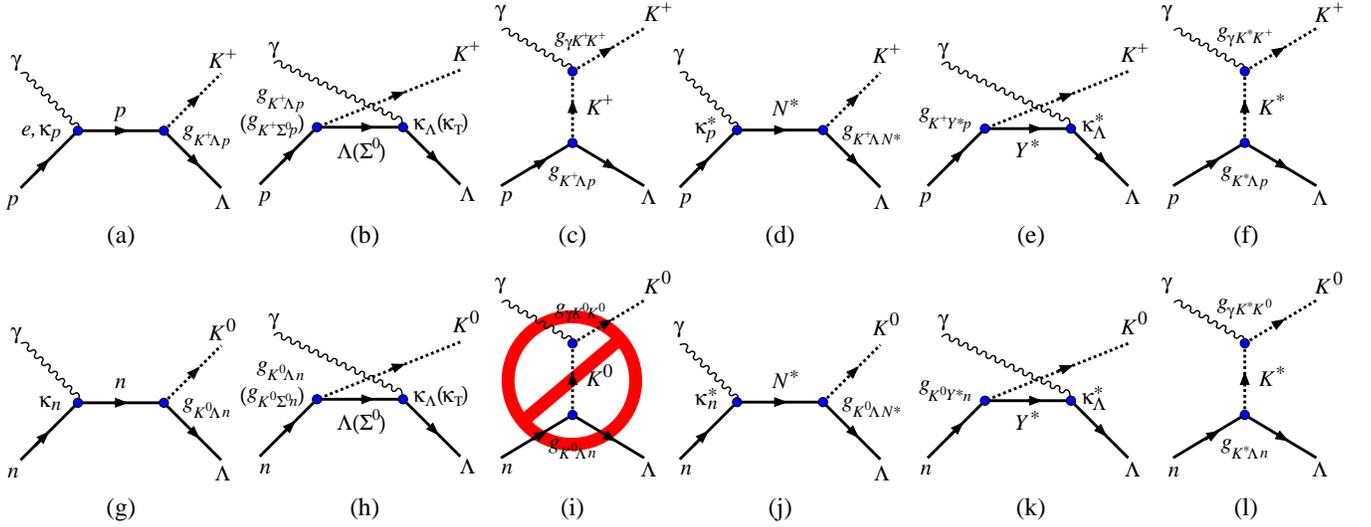}
\caption{Feynman diagrams for the $\gamma p\to K^+\Lambda$ 
         [(a)$-$(f)] and the $\gamma n\to K^0 \Lambda$ [(g)$-$(l)] photoproductions. 
         For the sake of brevity the contact diagrams for
         preserving the gauge invariance in the pseudovector theory, or in the
         pseudoscalar theory after including the hadronic form factors,
         are not displayed in this figure.}
\label{fig:feynman}
\end{figure*}

The reaction kinematics is defined through the general 
$K\Lambda$ photoproduction process,
\begin{eqnarray}
  \label{eq:reaction}
  \gamma (k) + N(p_N) \to K(q) + \Lambda(p_\Lambda) ~.
\end{eqnarray}
The relevant Feynman diagrams for the two processes considered in
this work are depicted in Fig.~\ref{fig:feynman}.
The  basic effective Lagrangian for the kaon-hyperon-nucleon interaction
can be written as
\begin{eqnarray}
  \label{eq:k-lam-N}
  {\cal L}_{K\Lambda N}=g_{K\Lambda N}\, {\bar \psi}_N \gamma_5
  \,\psi_{\Lambda}\,\Phi_K ,
\end{eqnarray}
and
\begin{eqnarray}
  \label{eq:k-sig-N}
  {\cal L}_{K\Sigma^0 N}=g_{K\Sigma^0 N}\, {\bar \psi}_N \gamma_5
  \,\psi_{\Sigma^0}\,\Phi_K ,
\end{eqnarray}
where $\psi_{N}$ and $\psi_\Lambda$ ($\psi_{\Sigma^0}$) are the spinor 
fields of nucleon and $\Lambda$ ($\Sigma^0$), respectively, and 
$\Phi_K$ is the pseudoscalar field of the kaon.

The electromagnetic interactions for the nucleon ($N$),
hyperon ($\Lambda$ and $\Sigma^0$), and kaon read
\begin{eqnarray}
  \label{eq:gam-NN}
  {\cal L}_{\gamma NN} &=& e {\bar \psi}_{N} \left( -Q_N \gamma_\mu A^\mu + 
  \frac{\kappa_{N}}{4m_N}
  \sigma_{\mu\nu}\, F^{\mu\nu} \right) \psi_N ,~~\\
  {\cal L}_{\gamma\Lambda\Lambda} &=& 
  \frac{e\kappa_{\Lambda}}{4m_\Lambda}
  {\bar \psi}_{\Lambda}
  \sigma_{\mu\nu}\, F^{\mu\nu} \psi_\Lambda , \\
  {\cal L}_{\gamma\Sigma^0\Lambda} &=& 
  \frac{e\kappa_{\rm T}}{2(m_\Lambda+m_{\Sigma^0})} {\bar \psi}_{\Lambda}
  \sigma_{\mu\nu}\, F^{\mu\nu} \psi_{\Sigma^0} ~,\\
  {\cal L}_{\gamma KK} &=& g_{\gamma KK} Q_K\left\{
  {\bar \Phi}_K (\partial_\mu \Phi_K) - (\partial_\mu {\bar \Phi_K}) 
  \Phi_K \right\} A^\mu ,~~
\end{eqnarray}
where $Q_N$ ($Q_K$) is the charge of the nucleon (kaon)
in the unit of positron charge, 
$F^{\mu\nu}=\partial^\mu A^\nu-\partial^\nu A^\mu$, while 
other terms are self-explanatory.

For the interactions of excited kaons the Lagrangians read 
\cite{Kim:2018qfu}
\begin{eqnarray}
  \label{eq:ks-lam-N}
  {\cal L}_{K^*\Lambda N} &=& {\bar \psi}_{N}\left\{ -i g^V_{K^*\Lambda N}
    \gamma^\mu\psi_\Lambda\right. \nonumber\\ && \left.
    + \frac{g^T_{K^*\Lambda N}}{m_N+m_\Lambda}
  \sigma_{\mu\nu}\,\psi_\Lambda \partial^{\nu} \right\} \Phi_\mu^{K^*} 
+ {\rm H.c.} ,
\end{eqnarray}
and 
\begin{eqnarray}
  \label{eq:gam-ks-k}
    {\cal L}_{\gamma KK^*} &=& g_{\gamma KK^*} 
    \epsilon_{\mu\nu\rho\sigma}\, \partial^\mu A^\nu 
       \nonumber\\ && \times 
       \left\{ 
      (\partial^\rho
        {\bar \Phi}^{\sigma}_{K^*}) \Phi_K 
        + {\bar \Phi}_K (\partial^\rho \Phi^{\sigma}_{K^*})
        \right\} ,~
\end{eqnarray}
where $\epsilon_{\mu \nu \rho \sigma}$ is the four-dimensional 
Levi-Civita tensor with
$\epsilon_{0123} = +1$. 

In the case of spin-1/2 nucleon resonances we also 
use the standard interaction Lagrangian,
i.e., \cite{Mart:2013ida}
\begin{eqnarray}
  \label{eq:hadronic_int}
  {\cal L}_{K\Lambda N^*}= g_{K\Lambda N^*} {\bar \psi}_\Lambda \gamma_5
  \,\psi_{N^*}\Phi_K + {\rm H.c.} ,
\end{eqnarray}
for the hadronic transition. For the magnetic transition the Lagrangian 
reads \cite{Mart:2013ida}
\begin{eqnarray}
  \label{eq:magnetic_int}
  {\cal L}_{\gamma NN^*}=\frac{eg_{\gamma NN^*}}{2(m_N+m_{N^*})}{\bar \psi}_{N^*}
  \sigma_{\mu\nu}\, F^{\mu\nu}\,\psi_N +{\rm H.c.} ,
\end{eqnarray}
where $g_{\gamma NN^*}$ is the transition magnetic moment
of the resonance. In our standard notation, the
propagator of spin-1/2 nucleon resonance reads
$(\slashed{p}_{N^*}+ 
m_{N^*})/(s-m_{N^*}^2+im_{N^*}\Gamma_{N^*})$, where 
$s$ is the Mandelstam variable and $\Gamma_{N^*}$
is the resonance width. Note that according to Eq.~(\ref{eq:reaction})
we have 
\begin{eqnarray}
  \label{eq:mandelstam_s}
  s &=& p_{N^*}^2 ~=~ (k+p_N)^2 ~=~ W^2 ~,\\
  \label{eq:mandelstam_t}
  t &=& (k-q)^2 ~,\\    
  \label{eq:mandelstam_u}
  u &=& (k-p_\Lambda)^2 ~.
\end{eqnarray}

For the nucleon resonances with spins 3/2 and higher we use the
consistent interaction Lagrangian described in Ref.~\cite{Clymton:2017nvp}.
In this case the hadronic interaction Lagrangian of the 
spin-($n+1/2$) nucleon resonance can be written as
\begin{eqnarray}
\mathcal{L}_{K\Lambda N^*}&=&\frac{g_{K\Lambda N^*}}{m_{N^*}^{2n+1}}\, 
\epsilon^{\mu\nu_n\alpha\beta}\,\partial^{\nu_1}\cdots
\partial^{\nu_{n-1}}\bar{\psi}_\Lambda \, \partial_{\beta} \Phi^{*}_K\, 
\gamma_5 \, \gamma_\alpha \, \nonumber \\
&&\times \partial_{\mu} \,\Psi_{\nu_1\cdots\nu_n}\, 
+ \mathrm{H.c.} ~,
\label{eq:lagrangian_had}
\end{eqnarray}
while the corresponding magnetic transition Lagrangian reads 
\begin{eqnarray}
\mathcal{L}_{\gamma NN^*} &=& \frac{e}{m_{N^*}^{2n+1}}\bar{\Psi}^{\beta_1\cdots
\beta_n}\bigl\{g_{\gamma NN^*}^{(1)}\,\epsilon_{\mu\nu\alpha\beta_n} \partial^\alpha 
\psi_N \nonumber \\ && 
+ g_{\gamma NN^*}^{(2)}\,\gamma_5 \, g_{\beta_n\nu}\partial_\mu \psi_N \nonumber \\ && 
+ g_{\gamma NN^*}^{(3)}\,\gamma_\mu\,\gamma^\rho\,\epsilon_{\rho\nu\alpha\beta_n}
\partial^\alpha \psi_N 
\nonumber\\ &&
+ g_{\gamma NN^*}^{(4)}\,\gamma_5\,\gamma_\mu\,\gamma^\rho\,
(\partial_\rho g_{\nu\beta_n} 
- \partial_\nu g_{\rho\beta_n}) \psi_N \bigr\}
 \nonumber\\&& 
\times\partial_{\beta_1} 
\cdots\partial_{\beta_{n-1}} F^{\mu\nu} + \mathrm{H.c.}
\label{eq:lagrangian_em}
\end{eqnarray}
Note that in Eqs.~(\ref{eq:lagrangian_had}) and (\ref{eq:lagrangian_em})
we have introduced $\Psi_{\mu_1\cdots\mu_n}$ to denote 
the massive Rarita-Schwinger field of $N^*$. 
Furthermore, Eq.~(\ref{eq:lagrangian_em})
also indicates that, in general, we have four transition moments
for the magnetic excitations of nucleon resonances with spins
3/2 or higher, which are indicated by $g_{\gamma NN^*}^{(1)},\cdots,g_{\gamma NN^*}^{(4)}$.
Due to the lack of information on these transition moments, 
for all nucleon resonances we can 
only extract the product of this moment and the hadronic coupling
constant, i.e.,  
\begin{eqnarray}
  \label{eq:nucl-res-coupling-constant}
  G_{N^*}^{(i)} &\equiv& G_{K\Lambda N^*}^{(i)} = g_{\gamma NN^*}^{(i)}\, g_{K\Lambda N^*} ~,
\end{eqnarray}
from fitting to experimental data. Note that 
Eq.~(\ref{eq:nucl-res-coupling-constant}),
but without index $i$, is also used for the spin-1/2 nucleon resonance
described by Eqs.~(\ref{eq:hadronic_int}) and (\ref{eq:magnetic_int}).

\subsection{Propagators}

The propagators for spin-$(n+1/2)$ nucleon resonances can be obtained 
from Ref.~\cite{Vrancx:2011qv}, i.e., 
\begin{eqnarray}
&&P^{\mu_1\cdots\mu_n,\nu_1\cdots\nu_n}_{(n+1/2)}(p_{N^*})\nonumber\\
&& =\, 
\frac{\slashed{p}_{N^*} + m_{N^*}}{p_{N^*}^2-m_{N^*}^2+im_{N^*}\Gamma_{N^*}}\,
\widetilde{\mathcal{P}}^{\mu_1\cdots\mu_n,\nu_1\cdots\nu_n}_{(n+1/2)}(p_{N^*}) , ~~~
\end{eqnarray}
where 
$\widetilde{\mathcal{P}}^{\mu_1\cdots\mu_n,\nu_1\cdots\nu_n}_{(n+1/2)}(p_{N^*})$
denotes the on-shell projection operator that 
is obtained from the off-shell projection operator 
${\mathcal{P}}^{\mu_1\cdots\mu_n,\nu_1\cdots\nu_n}_{(n+1/2)}(p_{N^*})$
with the replacements 
$\slashed{p}_{N^*} \to m_{N^*}$ and ${p}_{N^*}^2 \to m_{N^*}^2$.
For example, the propagator for spin-7/2 nucleon resonance reads
\cite{Clymton:2017nvp}
\begin{eqnarray}
\label{eq:propagator-7/2}
\mathop{P^{7/2}_{\mu\mu_1\mu_2}}_{\;\;\nu\nu_1\nu_2}=
\frac{s^3}{m_{N^*}^6}\frac{(\slashed{p}_{N^*}+
m_{N^*})}{(s-m_{N^*}^2+im_{N^*}\Gamma_{N^*})}
\mathop{\mathcal{P}^{7/2}_{\mu\mu_1\mu_2}}_{\;\;\nu\nu_1\nu_2}~,
\end{eqnarray}
where the spin-7/2 projection operator is given by 
\cite{Clymton:2017nvp}
\begin{eqnarray}
\label{eq:proj7}
\mathop{\mathcal{P}^{7/2}_{\mu_1\mu_2\mu_3}}_{\;\;\nu_1\nu_2\nu_3}
&=& \frac{1}{36}\sum_{\mathrm{P}(\mu),\mathrm{P}(\nu)}\Bigl\{ P_{\mu_1\nu_1}P_{\mu_2\nu_2}P_{\mu_3\nu_3}
 \nonumber\\ && 
-\textstyle{\frac{3}{7}}P_{\mu_1\mu_2}P_{\nu_1\nu_2}P_{\mu_3\nu_3}
 \nonumber \\&&
+\textstyle{\frac{3}{7}}\gamma^{\,\rho}\gamma^\sigma P_{\mu_1\rho}P_{\nu_1\sigma}P_{\mu_2\nu_2}P_{\mu_3\nu_3}\nonumber\\
  & & -\textstyle{\frac{3}{35}}\gamma^{\,\rho}\gamma^\sigma P_{\mu_1\rho}P_{\nu_1\sigma}P_{\mu_2\mu_3}P_{\nu_2\nu_3}\Bigr\}~,
\end{eqnarray}
where $\mathrm{P}(\mu)$ and $\mathrm{P}(\nu)$ indicate the permutations 
of all possible $\mu$ and $\nu$ indices, respectively, while 
$P_{\mu\nu} = -g_{\mu\nu}+p_{N^*\mu} p_{N^*\nu}/s$. 

The interaction Lagrangians described above are designed for 
positive parity intermediate states. For the negative parity 
states we have to modify these Lagrangians as explained in 
Ref.~\cite{Clymton:2017nvp}.

\subsection{Differences between  $K^+\Lambda$ and $K^0\Lambda$ 
photoproductions}
The appropriate Feynman diagrams for both $K^+\Lambda$ and 
$K^0\Lambda$ channels are depicted in Fig.~\ref{fig:feynman},
where the corresponding Born terms are displayed by the 
diagrams (a)$-$(c) and (g) and (h) of Fig.~\ref{fig:feynman}, respectively.
From Fig.~\ref{fig:feynman} it is clear that 
the difference between the two processes originates from the charge
of participating nucleon and kaon. Therefore, 
the hadronic coupling constants in both reactions can be related
by using isospin symmetry. Since the $\Lambda$ hyperon 
is  an SU(3) isosinglet, the symmetry prescribes that the corresponding 
$N\to K\Lambda$ 
hadronic couplings in both channels are equal, i.e.,
\begin{equation}
\label{eq:lambda_coupling}
g_{K^{+} \Lambda p} = g_{K^{0} \Lambda n} ~.
\end{equation}

As shown by the diagrams (b) and (h) in Fig.~\ref{fig:feynman}, 
the $\Sigma^0$ exchange 
is also allowed in the $u$-channel. However, different from $\Lambda$ 
hyperon, $\Sigma$ is an SU(3) isotriplet
and as a consequence we obtain \cite{Mart:1995wu}
\begin{equation}
\label{eq:sigma_coupling}
g_{K^{+} \Sigma^0 p} = -g_{K^{0} \Sigma^0 n} .~
\end{equation}

In the $N\to K^*\Lambda$ vertex the isospin symmetry also leads
to a similar relation to that of Eq.~(\ref{eq:lambda_coupling}), i.e.,
\begin{equation}
\label{eq:vector_coupling}
g^{V,T}_{K^{*+} \Lambda p} = g^{V,T}_{K^{*0} \Lambda n} .
\end{equation}
Equation~(\ref{eq:vector_coupling}) is also valid for the 
 vector meson $K_1(1270)$.

Similar to Eqs.~(\ref{eq:lambda_coupling}) and (\ref{eq:sigma_coupling}) 
the application of isospin symmetry in 
the $N^*\to K\Lambda$ and $N\to KY^*$ vertices results in
[diagrams (d) and (j)]
\begin{equation}
\label{eq:nucleon_res_coupling}
g_{K^{+} \Lambda N^*} = g_{K^{0} \Lambda N^*} ~,
\end{equation}
and with $Y^*=\Lambda^*, \Sigma^{0*}$ 
[diagrams (e) and (k)]
\begin{equation}
\label{eq:hyperon_res_coupling}
g_{K^{+} \Lambda^* p} = g_{K^{0} \Lambda^* n} ~~~,~~~
g_{K^{+} \Sigma^{0*} p} = -g_{K^{0} \Sigma^{0*} n} ~.
\end{equation}

The electromagnetic vertices for the proton, neutron, $\Lambda$, and 
$\Sigma^0$ exchanges shown by the diagrams (a) and (b)  in Fig.~\ref{fig:feynman}
are elementary and do not need explanation. The $K^+$ and $K^0$ intermediate states 
as given by the diagrams (c) and (i) are 
also the same. Since the real 
photon cannot interact with a neutral kaon, 
the $K^0$ intermediate state is not allowed in the amplitude of the
$K^0\Lambda$ photoproduction, in contrast to the case of 
$K^+\Lambda$ photoproduction.

In the case of spin-1/2 
nucleon resonances the transition moments in the two
channels [diagrams (d) and (j)] can be related to the corresponding 
resonance helicity amplitudes $A_{1/2}$, i.e., \cite{Mart:1995wu}
\begin{eqnarray}
  \label{eq:nucl-res-trans-moment}
  r_{N^*} \equiv \frac{g_{\gamma nN^*}}{g_{\gamma pN^*}} = \frac{A_{1/2}^n}{A_{1/2}^p} ~.
\end{eqnarray}
For the nucleon resonances with spins 3/2 and higher the corresponding
relations are more complicated \cite{Clymton:2017nvp}. 
Furthermore, the corresponding values of neutron helicity  
amplitudes $A_{1/2}^n$ and $A_{3/2}^n$ are mostly unknown. Therefore,
in this work these values are considered as free parameters during the
fitting process and for this purpose we define the ratio 
[see Eq.~(\ref{eq:lagrangian_em})]
\begin{eqnarray}
  \label{eq:nucl-res-trans-moment1}
  r_{N^*}^{(i)} \equiv \frac{g_{\gamma nN^*}^{(i)}}{g_{\gamma pN^*}^{(i)}} ~.
\end{eqnarray}
For hyperon resonances fortunately the transition moments in the two
channels are identical, as shown by the diagrams (e) and (k) in 
Fig.~\ref{fig:feynman}. 

Finally, the kaon resonance transition moments shown by the diagrams 
(f) and (l) in Fig.~\ref{fig:feynman} can be related to its decay width.
The result for $K^*(892)$ is \cite{Mart:1995wu}
\begin{eqnarray}
r_{K^0} \equiv 
g_{K^{*0} K^{0} \gamma}/g_{K^{*+} K^{+} \gamma} = -1.53\pm 0.20 \,.
\end{eqnarray}
For the vector meson $K_1(1270)$ the corresponding value is not
available and, therefore, we consider it also as a fitting parameter, i.e.,
\begin{eqnarray}
r_{K_1} \equiv g_{K^{0}_1 K^{0} \gamma}/g_{K^{+}_1 K^{+} \gamma}  \,.
\label{eq:K1_coupling}
\end{eqnarray}

\subsection{Nucleon and hyperon resonances used in the present analysis}

The number of nucleon resonances used in our analysis is limited by the
energy range of experimental data in our database. As in the previous work
\cite{Clymton:2017nvp} we further constrain the number by excluding the 
resonances with one-star rating in the overall status given by PDG. The
result is listed in Table~\ref{tab:partial-width} along with their properties
obtained from the PDG estimates \cite{pdg}. There are 18 nucleon resonances
involved in this analysis with spins up to 9/2. Since the estimated mass and width of each 
resonance have error bars, during the fitting process we vary their values 
within these error bars. Note that, in contrast to the pion and $K\Sigma$ 
photoproductions, delta resonances are not allowed in the $K^+\Lambda$ 
photoproduction due to the isospin conservation. 

\begin{table}[t]
\caption{Properties of the nucleon resonances used in the present analysis.
  Data are taken from the Review of Particle Properties of the PDG \cite{pdg}.}
\label{tab:partial-width}
\begin{tabular}{l c c c r}
\hline\hline\\[-2.5ex]
Resonance & ~~~$J^P$~~~ & $M$ (MeV) & ~~~$\Gamma$ (MeV)~~~ & Status \\[0.5ex]
\hline\\[-2.5ex]
$N(1440)P_{11}$ & $1/2^+$ & 1410 to 1470 & 250 to 450  & ****\\
$N(1520)D_{13}$ & $3/2^-$ & 1510 to 1520 & 100 to 120  & ****\\
$N(1535)S_{11}$ & $1/2^-$ & 1515 to 1545 & 125 to 175  & ****\\
$N(1650)S_{11}$ & $1/2^-$ & 1645 to 1670 & $104\pm 10$ & ****\\
$N(1675)D_{15}$ & $5/2^-$ & 1670 to 1680 & $120\pm 15$ & ****\\
$N(1680)F_{15}$ & $5/2^+$ & 1680 to 1690 & $118\pm 6 $ & *** \\
$N(1700)D_{13}$ & $3/2^-$ & 1650 to 1750 & 100 to 250  & *** \\
$N(1710)P_{11}$ & $1/2^+$ & 1680 to 1740 & 80 to 200   & ****\\
$N(1720)P_{13}$ & $3/2^+$ & 1680 to 1750 & 150 to 400  & ****\\
$N(1860)F_{15}$ & $5/2^+$ & $1860^{+100}_{-40}$& $270^{+140}_{-50}$ & ** \\
$N(1875)D_{13}$ & $3/2^-$ & 1850 to 1920 & 120 to 250 & ***\\
$N(1880)P_{11}$ & $1/2^+$ & 1830 to 1930 & 200 to 400  & ***  \\
$N(1895)S_{11}$ & $1/2^-$ & 1870 to 1920 & 80 to 200   & **** \\
$N(1900)P_{13}$ & $3/2^+$ & 1890 to 1950 & 100 to 320  & **** \\
$N(1990)F_{17}$ & $7/2^+$ & 1950 to 2100 & 200 to 400  & ** \\
$N(2000)F_{15}$ & $5/2^+$ & $2060\pm 30$& $390\pm 55$  & ** \\
$N(2060)D_{15}$ & $5/2^-$ & 2030 to 2200 & 300 to 450  & ***\\
$N(2120)D_{13}$ & $3/2^-$ & 2060 to 2160 & 260 to 360  & ***\\
$N(2190)G_{17}$ & $7/2^-$ & 2140 to 2220 & 300 to 500  & ****\\
$N(2220)H_{19}$ & $9/2^+$ & 2200 to 2300 & 350 to 500  & ****\\
$N(2250)G_{19}$ & $9/2^-$ & 2250 to 2320 & 300 to 600  & ****\\[0.5ex]
\hline\hline
\end{tabular}
\end{table}

\begin{table}[t]
\caption{Properties of the hyperon resonances used in the present analysis.
  Data are taken from the Review of Particle Properties of the PDG \cite{pdg}.}
\label{tab:partial-width-hyp}
\begin{tabular}{l c c c r}
\hline\hline\\[-2.5ex]
Resonance & ~~~$J^P$~~~ & $M$ (MeV) & ~~~$\Gamma$ (MeV)~~~ & Status \\[0.5ex]
\hline\\[-2.5ex]
$\Lambda(1405)S_{01}$ & $1/2^-$ & $1405.1^{+1.3}_{-1.0}$ & $50.5\pm 2.0$  & ****\\
$\Lambda(1520)D_{03}$ & $3/2^-$ & $1517^{+4}_{-4}$ & $15^{+10}_{-8}$  & ****\\
$\Lambda(1600)P_{01}$ & $1/2^+$ & $1544^{+3}_{-3}$ & $112^{+12}_{-2}$  & ***\\
$\Lambda(1670)S_{01}$ & $1/2^-$ & $1669^{+3}_{-8}$ & $19^{+18}_{-2}$  & ****\\
$\Lambda(1690)D_{03}$ & $3/2^-$ & $1697^{+6}_{-6}$ & $65\pm 14$  & ****\\
$\Lambda(1800)S_{01}$ & $1/2^-$ & 1720 to 1850 & 200 to 400  & *** \\
$\Lambda(1810)P_{01}$ & $1/2^+$ & 1750 to 1850 & 50 to 250  & *** \\
$\Lambda(1890)P_{03}$ & $3/2^+$ & 1850 to 1910 & 60 to 200   & ****\\
$\Sigma(1385)P_{13}$  & $3/2^+$ & $1382.80\pm 0.35$ & $36.0\pm 0.7$  & ****\\
$\Sigma(1660)P_{11}$  & $1/2^+$ & 1630 to 1690 & 40 to 200  & *** \\
$\Sigma(1670)D_{13}$  & $3/2^-$ & 1665 to 1685 & 40 to 80 & ****\\
$\Sigma(1750)S_{11}$  & $1/2^-$ & 1730 to 1800 & 60 to 160  & ***  \\
$\Sigma(1880)P_{11}$  & $1/2^+$ & 1880 & $300\pm 59$   & ** \\
$\Sigma(1940)D_{13}$  & $3/2^-$ & $1941\pm 18$ & $400\pm 49$  & * \\
$\Sigma(2080)P_{13}$  & $3/2^+$ & 2080 & $186\pm 48$  & ** \\[0.5ex]
\hline\hline
\end{tabular}
\end{table}

For the hyperon resonances we limit the number of the used resonances
by excluding those with spins higher than 3/2. They are listed in
Table \ref{tab:partial-width-hyp}. Note that in the literatures
these resonances are considered as a part of the background, since
their squared momentum is $u$, instead of $s$.

\subsection{Experimental data and fitted observables}

\begin{table*}[t]
 \centering
  \caption{Observables and experimental data used in the previous and present studies. Channels
    1 and 2 refer to the $\gamma p \to K^+\Lambda$ and $\gamma n\to K^0\Lambda$
    channels, respectively.
    Columns M1, M2, and M3 indicate the data set used in the models M1, M2, and M3, 
    respectively. The data set used in the previous model \cite{Clymton:2017nvp} 
    and updated to include the CLAS 2016 double polarization data \cite{paterson}
    is indicated by M0 \cite{samson_thesis}. Note that, except for channel 2, all data sets listed in this 
    table have been given in our previous work  \cite{Mart:2017mwj}.}
  \label{tab:experimental_data}
  \begin{ruledtabular}
    \begin{tabular}[c]{llcrcccccc}
      Collaboration & Observable & Symbol & $N$~ & Channel& M0 & M1 & M2 & M3 & Reference  \\
      \hline
      CLAS 2006 & Differential cross section & $d\sigma/d\Omega$ & 1377 & 1 & \checkmark & \checkmark & \checkmark & \checkmark &\cite{Bradford:2005pt} \\
      & Recoil polarization & $P$ & 233                                 & 1 & \checkmark & \checkmark & \checkmark & \checkmark &\cite{Bradford:2005pt} \\
      LEPS 2006 & Differential cross section & $d\sigma/d\Omega$ & $54$ & 1 & \checkmark & \checkmark & \checkmark & \checkmark & \cite{Sumihama06} \\
      & Photon asymmetry & $\Sigma$ & $30$                              & 1 & \checkmark & \checkmark & \checkmark & \checkmark & \cite{Sumihama06}\\
      GRAAL 2007& Recoil polarization & $P$ & $66$                      & 1 & \checkmark & \checkmark & \checkmark & \checkmark & \cite{lleres:2007} \\
      & Photon asymmetry & $\Sigma$ & $66$                              & 1 & \checkmark & \checkmark & \checkmark & \checkmark & \cite{lleres:2007}\\
      LEPS 2007 & Differential cross section & $d\sigma/d\Omega$ & $12$ & 1 & \checkmark & \checkmark & \checkmark & \checkmark & \cite{Hicks_2007}\\
      CLAS 2007 & Beam-Recoil polarization & $C_x$ & $159$              & 1 & \checkmark & \checkmark & \checkmark & \checkmark & \cite{Bradford:2006ba}\\
      & Beam-Recoil polarization & $C_z$ & 160                          & 1 & \checkmark & \checkmark & \checkmark & \checkmark & \cite{Bradford:2006ba} \\
      GRAAL 2009& Target asymmetry & $\Sigma$ & $66$                    & 1 & \checkmark & \checkmark & \checkmark & \checkmark & \cite{lleres:2009}\\
      & Beam-Recoil polarization & $O_{x'}$ & $66$                      & 1 & \checkmark & \checkmark & \checkmark & \checkmark & \cite{lleres:2009} \\
      & Beam-Recoil polarization & $O_{z'}$ & $66$                      & 1 & \checkmark & \checkmark & \checkmark & \checkmark & \cite{lleres:2009} \\
      CLAS 2010 & Differential cross section & $d\sigma/d\Omega$ & 2066 & 1 & \checkmark & \checkmark & \checkmark & \checkmark & \cite{mcCracken} \\
      & Recoil polarization & $P$ & 1707                                & 1 & \checkmark & \checkmark & \checkmark & \checkmark & \cite{mcCracken} \\
      Crystal Ball 2014&                                                                                       
      Differential cross section & $d\sigma/d\Omega$  & 1301            & 1 & \checkmark & \checkmark & \checkmark & \checkmark & \cite{Jude:2013jzs} \\
      CLAS 2016 & Recoil polarization & $P$ & $314$                     & 1 & \checkmark & \checkmark & \checkmark & \checkmark & \cite{paterson} \\
      & Photon asymmetry & $\Sigma$ &  $314$                            & 1 & \checkmark & \checkmark & \checkmark & \checkmark & \cite{paterson} \\
      & Target asymmetry & $T$ & $314$                                  & 1 & \checkmark & \checkmark & \checkmark & \checkmark & \cite{paterson}\\
      & Beam-Recoil polarization & $O_x$ & $314$                        & 1 & \checkmark & \checkmark & \checkmark & \checkmark & \cite{paterson} \\
      & Beam-Recoil polarization & $O_z$ & $314$                        & 1 & \checkmark & \checkmark & \checkmark & \checkmark & \cite{paterson} \\
      CLAS 2017 & Differential cross section & $d\sigma/d\Omega$ & $361$& 2 & $\cdots$   & \checkmark & $\cdots$   & \checkmark &\cite{Compton:2017xkt} \\
     MAMI 2018 & Differential cross section & $d\sigma/d\Omega$ & $60$ & 2 & $\cdots$   & $\cdots$   & \checkmark & \checkmark & \cite{Akondi:2018shh} \\[1ex]
      \hline                                                              
      \multicolumn{3}{l}{Total number of data }                  & 9424 &   & $9003 $    & $9364$     & $9063$     &$9424$      & \\
    \end{tabular}
  \end{ruledtabular}
\end{table*}

As explained in the introduction the present model is based on our previous 
covariant isobar one \cite{Clymton:2017nvp} that fits nearly 7400 
experimental data points of the $\gamma p \to K^+\Lambda$ channel. However,
we notice that additional experimental data for double polarization observables
\cite{paterson}
consisting of 1574 data points appeared almost at the same time of the 
publication of Ref.~\cite{Clymton:2017nvp}. Thus, in total the number of 
$\gamma p \to K^+\Lambda$ data points is 9003. At this stage it is important
to mention that a new determination of $\Lambda$ decay parameter $\alpha_-$
has been reported in Ref.~\cite{Ireland:2019uja}. The new value is significantly
larger than the standard one tabulated in the Review of Particle Properties of 
PDG \cite{pdg} and may affect the obtained experimental data for the recoil 
polarization. Note that, different from 
our previous covariant isobar model, our recent multipole analysis 
\cite{Mart:2017mwj} has used 9003 data points. As a consequence,
we have to refit our previous covariant model by including these double 
polarization data in the fitting database. The result is denoted by M0 and 
is discussed in Sec.~\ref{sec:result}. The detailed data sets used in 
this model are listed in Table~\ref{tab:experimental_data}.

As shown in Table~\ref{tab:experimental_data} the new $\gamma n\to K^0 \Lambda$ 
data consist of 401 data points obtained from the CLAS 2017 \cite{Compton:2017xkt} 
and MAMI 2018 \cite{Akondi:2018shh} collaborations. Our recent analysis on this
isospin channel indicates
that the two data sets show a sizable discrepancy \cite{Mart:ptep2019}. To investigate 
the effect of these two data sets on our model we propose three different models that
fit all $\gamma p \to K^+\Lambda$ data along with the $\gamma n\to K^0 \Lambda$ data
obtained from the CLAS 2017, MAMI 2018, and both CLAS 2017 and MAMI 2018 
collaborations. In Table~\ref{tab:experimental_data} the three different models are 
denoted by M1, M2, and M3, respectively.

\subsection{Calculation of observables}
For the convenience of the reader we briefly summarize the formulas used for calculating
the observables. 
Since we are working with photoproduction the formalism is further simplified
by the fact that $k^2=k\cdot\epsilon=0$. The transition amplitude ${\cal M}$ obtained from
the Feynman diagrams given in Fig.~\ref{fig:feynman} can be decomposed into 
the gauge and Lorentz invariant matrices $M_{i}$ \cite{Clymton:2017nvp},
\begin{eqnarray}
{\cal M} &=& {\bar u}_\Lambda \sum_{i=1}^4 A_{i}(s,t,u)\, M_{i}\, u_N~ ,
\label{eq:scattering-amplitudes-Mi}
\end{eqnarray}
where $s$, $t$, and $u$ are the Mandelstam variables given in
Eqs.~(\ref{eq:mandelstam_s})$-$(\ref{eq:mandelstam_u}) and 
the four gauge and Lorentz invariant matrices $M_i$  
are given by 
\begin{eqnarray}
\label{eq:M1}
M_{1} & = & \gamma_{5} \, \epsilon\!\!/ k\!\!\!/ ~ ,\\
\label{eq:M2}
M_{2} & = & 2\gamma_{5}\left(q \cdot \epsilon P \cdot k - q
\cdot k P \cdot \epsilon \right)~ ,\\
\label{eq:M3}
M_{3} & = & \gamma_{5} \left( q\cdot k \epsilon\!\!/ - q\cdot \epsilon 
k\!\!\!/ \right) ~ ,\\
\label{eq:M4}
M_{4} & = & i \epsilon_{\mu \nu \rho \sigma} \gamma^{\mu} q^{\nu}
\epsilon^{\rho} k^{\sigma}~ ,
\end{eqnarray}
with $P = \frac{1}{2}(p_N + p_{\Lambda})$ and 
$\epsilon$ being the photon polarization.
All observables required in the present analysis can be calculated
from $A_i$ extracted from Eq.~(\ref{eq:scattering-amplitudes-Mi}).
The relations between these observables and $A_i$ 
can be found, e.g., in Ref.~\cite{Knochlein:1995qz}.

In the hadronic vertices of background and resonance terms 
we insert hadronic form factors in the form of  \cite{Haberzettl:1998eq}
\begin{eqnarray}
  \label{eq:hadr_ff}
  F(\Lambda,x)&=& \frac{\Lambda^4}{\Lambda^4+(x-m_x^2)^2} ~,
\end{eqnarray}
where $x$, $\Lambda$, and $m_x$ are the corresponding Mandelstam variables 
($s$, $t$, or $u$), the form factor cutoff, and the mass of
intermediate state, respectively.
Note that in the background terms the inclusion of hadronic form factors
destroys the gauge invariance of the amplitude. To restore the gauge invariance
we utilize the Haberzettl method \cite{Haberzettl:1998eq} with the cost of adding 
two extra parameters originating from the freedom of choosing the form factor 
for the electric amplitude $A_2$ in Eq.~(\ref{eq:scattering-amplitudes-Mi}). 
As in the previous study we use \cite{Haberzettl:1998eq}
\begin{eqnarray}
{\tilde F}(\Lambda,s, t, u) &=& F(\Lambda,s)  \sin^2\theta_{\rm had} \cos^2\phi_{\rm had}
\nonumber \\ && +
        F(\Lambda,u) \sin^2\theta_{\rm had}\sin^2\phi_{\rm had}
\nonumber \\ && + F(\Lambda,t) 
        \cos^2\theta_{\rm had} ~,
\label{eq:fhabtilde}
\end{eqnarray}
where the combination of the form factors in Eq.~(\ref{eq:fhabtilde})
ensures the correct normalization,
i.e., ${\tilde F}(\Lambda,m_s^2, m_t^2, m_u^2)=1$. In the present work we
extract both $\theta_{\rm had}$ and $\phi_{\rm had}$ from the fitting process.
Our previous investigations indicate that the background and resonance amplitudes 
require different suppressions from the form factors. Therefore, in the present
work we separate
these form factors by defining different cutoffs for these amplitudes, i.e., 
$\Lambda_{\rm B}$ and $\Lambda_{\rm R}$, respectively.

\section{RESULTS AND DISCUSSION}
\label{sec:result}

\subsection{Results from previous works}
\begin{figure}
  \includegraphics[scale=0.372]{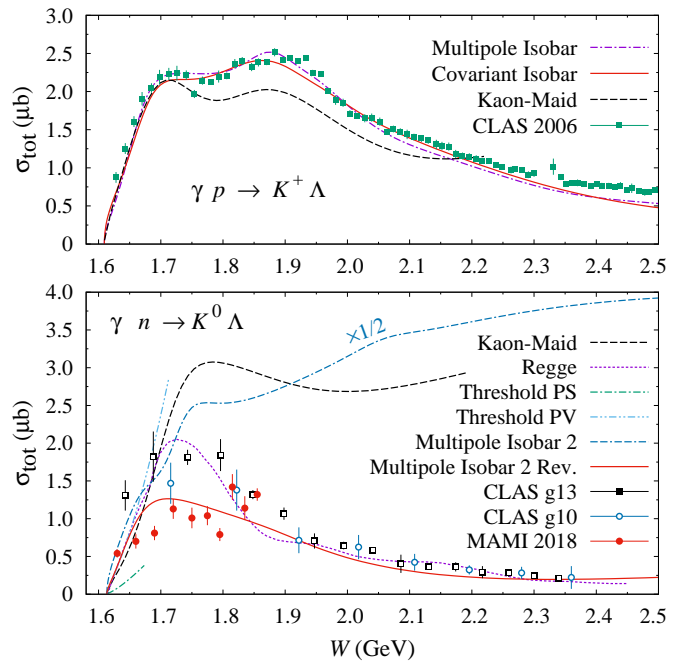}
\caption{(Top panel) The $K^+\Lambda$ photoproduction total cross section 
  as a function of the total c.m. energy $W$. Experimental data are taken
  from the CLAS 2006 collaboration \cite{Bradford:2005pt}. The dot-dashed
  curve shows the result of the recent multipole analysis of the
  $\gamma p\to K^+\Lambda$ photoproduction data \cite{Mart:2017mwj},
  and the dashed curve is taken from the Kaon-Maid model  \cite{kaon-maid}, 
  whereas the solid curve is obtained from the previous effective Lagrangian 
  model with nucleon spins up to 9/2 \cite{Clymton:2017nvp}. 
  (Bottom panel) As in the top panel but for 
  the $\gamma n\to K^0\Lambda$ photoproduction. Experimental data are 
  taken from the CLAS g10 and g13 experiments \cite{Compton:2017xkt},
  as well as from the MAMI 2018 collaboration \cite{Akondi:2018shh}.
  The dashed and dotted curves show the result of Kaon-Maid \cite{kaon-maid}
  and recent effective-Lagrangian-Regge \cite{Kim:2018qfu} models,
  respectively. The dot-dashed and dot-dot-dashed curves are obtained from the
  photoproduction analysis at threshold by using PS and
  PV couplings \cite{mart_thr}, respectively, whereas the
  the dot-dash-dashed and solid curves exhibit the prediction of the 
  recent multipole model \cite{Mart:2017xtf} and the corresponding 
  revised version \cite{Mart:ptep2019}. Note that the dot-dash-dashed curve
  has been rescaled by a factor of 1/2 to fit in the same plot. All data shown
  in this figure were not used in the fitting process; they are shown here only
  for comparison.}
\label{fig:tot_prob}
\end{figure}

The results from our previous analyses are shown in Fig.~\ref{fig:tot_prob}.
For the $\gamma p\to K^+\Lambda$ channel, except in the case of Kaon-Maid, 
it is clear that our previous calculations can nicely reproduce the 
experimental data.  As shown in the top panel of Fig.~\ref{fig:tot_prob}
Kaon-Maid cannot reproduce the data because it was fitted to the SAPHIR 
data \cite{Tran:1998,Glander:2003jw} which are smaller than the CLAS data for 
$W\geq 1.7$ GeV. This problem has been thoroughly discussed in
Ref.~\cite{Mart:2006dk}. For $W\geq 2.3$ GeV both multipole and isobar models 
seem to underpredict the CLAS total cross section data. However, this problem does not 
appear in most of the differential cross section data used in the fitting database because 
they are in fact smaller than the CLAS 2006 ones (see Fig.~\ref{fig:dkple} and the corresponding
discussion below). Thus, we believe that the smaller calculated 
total cross section shown in this kinematics is natural

In contrast to the  $\gamma p\to K^+\Lambda$ channel, as shown in the
bottom panel of Fig.~\ref{fig:tot_prob} the predicted and calculated
$\gamma n\to K^0\Lambda$ total cross sections vary significantly. This is
understandable because without fitting to the experimental data the isospin 
symmetry relations given by Eqs.~(\ref{eq:lambda_coupling})-(\ref{eq:K1_coupling})
could lead to a large transition amplitude that eventually yields a huge 
cross section. We note that this situation happens in the Kaon-Maid and 
our previous~\cite{Mart:2017xtf}  models. However, in the latter it has been shown that 
the model can be 
significantly improved by including the $\gamma n\to K^0\Lambda$ differential 
cross section data in the fitting database~\cite{Mart:ptep2019}, although 
as shown by the solid line in the bottom panel of Fig.~\ref{fig:tot_prob} 
this model exhibits a compromise result since it was fitted to both CLAS and
MAMI data.

In spite of the wild variation of the models shown in the bottom panel of 
Fig.~\ref{fig:tot_prob}, within the experimental error bars all models are 
still in agreement with the trend of experimental data in the low energy region, 
i.e., from threshold up to $W\approx 1.66$ GeV. Especially interesting is the 
result of our previous investigation that explores the difference between the 
use of PS and PV couplings in the threshold 
region \cite{mart_k0lambda}. Although
it was believed that the PS model works better than the PV one, comparison between
the two models given in the bottom panel of Fig.~\ref{fig:tot_prob} indicates that
PV coupling yields a better agreement with the data in the low energy region. The PS coupling
yields a much smaller total cross section; still smaller than the data given by
the MAMI 2018 collaboration. Nevertheless, a firm conclusion to this end cannot
be easily drawn before we can solve the problem of discrepancy between the CLAS
and MAMI data. We discuss this problem in the following subsection.

\subsection{Results from the present analysis}

\begin{table}[!]
 \centering
  \caption{Coupling constants and some parameters of the background terms.
  See Ref.~\cite{Mart:2015jof} for the notation of the parameters.}
  \label{tab:background}
  \begin{ruledtabular}
  \centering
  \begin{tabular}{lrrrr} 
    Parameter & M0 & M1 & M2 & M3 \\[1ex]
    \hline\\[-2.2ex]
  $g_{K \Lambda N}/\sqrt{4\pi}$  & $ -3.43$ & $ -3.00$ & $ -3.96$ & $ -3.40$ \\    
  $g_{K \Sigma N}/\sqrt{4\pi}$   & $  1.30$ & $  1.30$ & $  1.30$ & $  1.30$ \\    
  $G^{V}_{K^{*}}/4\pi$           & $  0.22$ & $  0.13$ & $  0.22$ & $  0.18$ \\    
  $G^{T}_{K^{*}}/4\pi$           & $  0.37$ & $  0.17$ & $  0.36$ & $  0.18$ \\    
  $G^{V}_{K_1}/4\pi$             & $ -0.07$ & $  0.13$ & $ -0.25$ & $ -0.17$ \\    
  $G^{T}_{K_1}/4\pi$             & $  4.40$ & $  3.89$ & $  4.54$ & $  4.48$ \\    
  $r_{K_1K\gamma}$             & $  0.00$ & $  0.65$ & $ -0.52$ & $  0.69$\\
  $\theta_{\rm had}$ (deg.)    & $  90.0$ & $  90.0$ & $  90.0$ & $  90.0$\\
  $\phi_{\rm had}$   (deg)     & $144.6 $ & $  0.0 $ & $ 60.6 $ & $  8.0 $\\
  $\Lambda_{\rm B}$ (GeV)      & $  0.70$ & $  0.70$ & $  0.70$ & $  0.70$ \\
  $\Lambda_{\rm R}$ (GeV)      & $  1.07$ & $  1.10$ & $  1.07$ & $  1.09$ \\
  \hline\\[-2.2ex]
  $\chi^2$                     &13433    & 13867 & 13473 & 14068 \\
  $N_{\rm data} $              & 9003    &  9364 &  9063 &  9424 \\
  $N_{\rm par.} $              &  184    &   247 &   247 &   247 \\
  $\chi^2/N_{\rm dof}$         & 1.52    &  1.52 &  1.53 &  1.53 \\[1ex]
    \end{tabular}
  \end{ruledtabular}
\end{table}

\begin{table*}[!]
 \centering
  \caption{Properties of the nucleon resonances extracted from different models.
    Note that $G^{(i)}_{N^*}$ and $r^{(i)}_{N^*}$, with $i=1,\cdots,4$, are defined in 
  Eqs.~(\ref{eq:nucl-res-coupling-constant}) and (\ref{eq:nucl-res-trans-moment1}),
  respectively. }
  \label{tab:mass_nuc_extracted}
  \begin{ruledtabular}
    \begin{tabular}[c]{lcccrrrrrrrr}
      Resonance & Model& $m_{N^*}$& $\Gamma_{N^*}$& $G^{(1)}_{N^*}$ & $G^{(2)}_{N^*}$ & $G^{(3)}_{N^*}$ & $G^{(4)}_{N^*}$ & $r^{(1)}_{N^*}$ & $r^{(2)}_{N^*}$ & $r^{(3)}_{N^*}$ & $r^{(4)}_{N^*}$   \\
      & &  (MeV) &  (MeV) & &&&&&&&\\
      \hline
$N(1440)P_{11}$ & M0 & $1410$ & $450$ & $  0.77$ & $\cdots$ & $\cdots$ & $\cdots$ & $\cdots$ & $\cdots$ & $\cdots$ & $\cdots$ \\
& M1 & $1410$ & $450$ & $  0.64$ & $\cdots$ & $\cdots$ & $\cdots$ & $\cdots$ & $\cdots$ & $\cdots$ & $\cdots$ \\
& M2 & $1410$ & $450$ & $  0.67$ & $\cdots$ & $\cdots$ & $\cdots$ & $\cdots$ & $\cdots$ & $\cdots$ & $\cdots$ \\
& M3 & $1410$ & $450$ & $  0.80$ & $\cdots$ & $\cdots$ & $\cdots$ & $\cdots$ & $\cdots$ & $\cdots$ & $\cdots$ \\
$N(1520)D_{13}$ & M0 & $1520$ & $125$ & $  0.12$ & $ -0.85$ & $ -0.09$ & $  0.25$ & $  0.00$ & $  0.00$ & $  0.00$ & $  0.00$ \\
& M1 & $1520$ & $100$ & $  0.18$ & $ -1.25$ & $ -0.01$ & $  0.41$ & $ -0.04$ & $  0.98$ & $  2.00$ & $  1.10$ \\
& M2 & $1520$ & $100$ & $  0.18$ & $ -1.43$ & $ -0.07$ & $  0.52$ & $  1.46$ & $ -0.21$ & $  1.64$ & $ -0.31$ \\
& M3 & $1520$ & $100$ & $  0.19$ & $ -1.62$ & $ -0.04$ & $  0.59$ & $  0.36$ & $  1.58$ & $ -1.04$ & $  1.64$ \\
$N(1535)S_{11}$ & M0 & $1525$ & $125$ & $ -0.11$ & $\cdots$ & $\cdots$ & $\cdots$ & $\cdots$ & $\cdots$ & $\cdots$ & $\cdots$ \\
& M1 & $1525$ & $125$ & $ -0.10$ & $\cdots$ & $\cdots$ & $\cdots$ & $\cdots$ & $\cdots$ & $\cdots$ & $\cdots$ \\
& M2 & $1545$ & $175$ & $  0.07$ & $\cdots$ & $\cdots$ & $\cdots$ & $\cdots$ & $\cdots$ & $\cdots$ & $\cdots$ \\
& M3 & $1545$ & $175$ & $  0.05$ & $\cdots$ & $\cdots$ & $\cdots$ & $\cdots$ & $\cdots$ & $\cdots$ & $\cdots$ \\
$N(1650)S_{11}$ & M0 & $1645$ & $150$ & $  0.15$ & $\cdots$ & $\cdots$ & $\cdots$ & $\cdots$ & $\cdots$ & $\cdots$ & $\cdots$ \\
& M1 & $1645$ & $155$ & $  0.15$ & $\cdots$ & $\cdots$ & $\cdots$ & $\cdots$ & $\cdots$ & $\cdots$ & $\cdots$ \\
& M2 & $1645$ & $170$ & $  0.09$ & $\cdots$ & $\cdots$ & $\cdots$ & $\cdots$ & $\cdots$ & $\cdots$ & $\cdots$ \\
& M3 & $1645$ & $159$ & $  0.09$ & $\cdots$ & $\cdots$ & $\cdots$ & $\cdots$ & $\cdots$ & $\cdots$ & $\cdots$ \\
$N(1675)D_{15}$ & M0 & $1680$ & $130$ & $ -2.87$ & $ -0.20$ & $ -2.43$ & $  0.18$ & $  0.00$ & $  0.00$ & $  0.00$ & $  0.00$ \\
& M1 & $1680$ & $130$ & $ -2.62$ & $ -0.08$ & $ -2.25$ & $  0.13$ & $  0.65$ & $ -2.00$ & $  0.76$ & $  1.63$ \\
& M2 & $1680$ & $130$ & $ -2.74$ & $ -0.35$ & $ -2.33$ & $  0.23$ & $  1.35$ & $ -2.00$ & $  1.19$ & $  2.00$ \\
& M3 & $1680$ & $130$ & $ -2.61$ & $ -0.33$ & $ -2.23$ & $  0.24$ & $ -0.48$ & $  1.78$ & $ -0.54$ & $  1.88$ \\
$N(1680)F_{15}$ & M0 & $1690$ & $120$ & $  0.93$ & $ 10.00$ & $  0.50$ & $ -4.84$ & $  0.00$ & $  0.00$ & $  0.00$ & $  0.00$ \\
& M1 & $1690$ & $120$ & $  1.02$ & $ 10.00$ & $  0.70$ & $ -4.77$ & $ -1.28$ & $  1.64$ & $ -1.42$ & $  1.41$ \\
& M2 & $1690$ & $120$ & $  1.10$ & $ 10.00$ & $  0.70$ & $ -4.74$ & $  2.00$ & $ -0.59$ & $ -2.00$ & $ -0.53$ \\
& M3 & $1690$ & $120$ & $  1.01$ & $ 10.00$ & $  0.62$ & $ -4.77$ & $ -1.61$ & $  0.05$ & $ -2.00$ & $  0.04$ \\
$N(1700)D_{13}$ & M0 & $1732$ & $123$ & $  0.13$ & $  0.84$ & $  0.29$ & $ -0.27$ & $  0.00$ & $  0.00$ & $  0.00$ & $  0.00$ \\
& M1 & $1735$ & $130$ & $  0.15$ & $  0.77$ & $  0.29$ & $ -0.24$ & $  2.00$ & $ -0.05$ & $  0.98$ & $  0.37$ \\
& M2 & $1731$ & $113$ & $  0.09$ & $  0.72$ & $  0.24$ & $ -0.23$ & $ -1.97$ & $ -0.74$ & $ -1.26$ & $ -1.16$ \\
& M3 & $1731$ & $102$ & $  0.06$ & $  0.62$ & $  0.19$ & $ -0.20$ & $ -2.00$ & $ -0.05$ & $ -0.69$ & $  0.04$ \\
$N(1710)P_{11}$ & M0 & $1740$ & $250$ & $  0.26$ & $\cdots$ & $\cdots$ & $\cdots$ & $\cdots$ & $\cdots$ & $\cdots$ & $\cdots$ \\
& M1 & $1740$ & $250$ & $  0.25$ & $\cdots$ & $\cdots$ & $\cdots$ & $\cdots$ & $\cdots$ & $\cdots$ & $\cdots$ \\
& M2 & $1729$ & $250$ & $  0.30$ & $\cdots$ & $\cdots$ & $\cdots$ & $\cdots$ & $\cdots$ & $\cdots$ & $\cdots$ \\
& M3 & $1733$ & $250$ & $  0.26$ & $\cdots$ & $\cdots$ & $\cdots$ & $\cdots$ & $\cdots$ & $\cdots$ & $\cdots$ \\
$N(1720)P_{13}$ & M0 & $1700$ & $154$ & $ -0.05$ & $  0.23$ & $ -0.02$ & $  0.00$ & $  0.00$ & $  0.00$ & $  0.00$ & $  0.00$ \\
& M1 & $1700$ & $159$ & $ -0.04$ & $  0.22$ & $ -0.01$ & $  0.01$ & $  0.12$ & $  1.34$ & $  2.00$ & $  2.00$ \\
& M2 & $1700$ & $171$ & $ -0.17$ & $  0.30$ & $ -0.11$ & $ -0.02$ & $  1.88$ & $  0.61$ & $  2.00$ & $ -0.99$ \\
& M3 & $1700$ & $189$ & $ -0.19$ & $  0.32$ & $ -0.13$ & $ -0.02$ & $  1.70$ & $  1.36$ & $  2.00$ & $ -2.00$ \\
$N(1860)F_{15}$ & M0 & $1960$ & $220$ & $ -0.99$ & $ -9.70$ & $ -0.18$ & $  4.25$ & $  0.00$ & $  0.00$ & $  0.00$ & $  0.00$ \\
& M1 & $1960$ & $220$ & $ -1.16$ & $ -8.80$ & $ -0.35$ & $  3.84$ & $ -0.01$ & $  1.72$ & $ -0.97$ & $  1.23$ \\
& M2 & $1960$ & $220$ & $ -1.07$ & $ -9.90$ & $ -0.32$ & $  4.24$ & $ -2.00$ & $  1.61$ & $ -2.00$ & $  0.44$ \\
& M3 & $1960$ & $220$ & $ -1.14$ & $ -9.01$ & $ -0.30$ & $  3.90$ & $  1.26$ & $ -2.00$ & $  1.90$ & $ -1.99$ \\
$N(1875)D_{13}$ & M0 & $1920$ & $320$ & $  0.06$ & $  0.40$ & $  0.09$ & $ -0.14$ & $  0.00$ & $  0.00$ & $  0.00$ & $  0.00$ \\
& M1 & $1889$ & $180$ & $  0.00$ & $  0.11$ & $  0.03$ & $ -0.03$ & $ -2.00$ & $  0.59$ & $ -2.00$ & $ -2.00$ \\
& M2 & $1873$ & $180$ & $ -0.02$ & $  0.27$ & $  0.03$ & $ -0.10$ & $  2.00$ & $ -0.75$ & $  2.00$ & $ -1.39$ \\
& M3 & $1858$ & $180$ & $ -0.03$ & $  0.40$ & $  0.01$ & $ -0.15$ & $ -2.00$ & $ -1.46$ & $  2.00$ & $ -2.00$ \\
    \end{tabular}
  \end{ruledtabular}
\end{table*}

\begin{table*}[!h]
  \addtocounter{table}{-1}
 \centering
  \caption{Properties of the nucleon resonances extracted from different models (continued).}
  \begin{ruledtabular}
    \begin{tabular}[c]{lcccrrrrrrrr}
      Resonance & Model& $m_{N^*}$& $\Gamma_{N^*}$& $G^{(1)}_{N^*}$ & $G^{(2)}_{N^*}$ & $G^{(3)}_{N^*}$ & $G^{(4)}_{N^*}$ & $r^{(1)}_{N^*}$ & $r^{(2)}_{N^*}$ & $r^{(3)}_{N^*}$ & $r^{(4)}_{N^*}$   \\
      & &  (MeV) &  (MeV) & &&&&&&&\\
      \hline
$N(1880)P_{11}$ & M0 & $1915$ & $280$ & $  0.17$ & $\cdots$ & $\cdots$ & $\cdots$ & $  0.00$ & $\cdots$ & $\cdots$ & $\cdots$ \\
& M1 & $1915$ & $280$ & $  0.19$ & $\cdots$ & $\cdots$ & $\cdots$ & $  0.96$ & $\cdots$ & $\cdots$ & $\cdots$ \\
& M2 & $1915$ & $280$ & $  0.18$ & $\cdots$ & $\cdots$ & $\cdots$ & $ -0.88$ & $\cdots$ & $\cdots$ & $\cdots$ \\
& M3 & $1915$ & $280$ & $  0.18$ & $\cdots$ & $\cdots$ & $\cdots$ & $  0.16$ & $\cdots$ & $\cdots$ & $\cdots$ \\
$N(1895)S_{11}$ & M0 & $1893$ & $107$ & $ -0.03$ & $\cdots$ & $\cdots$ & $\cdots$ & $  0.00$ & $\cdots$ & $\cdots$ & $\cdots$ \\
& M1 & $1893$ & $106$ & $ -0.03$ & $\cdots$ & $\cdots$ & $\cdots$ & $ -0.39$ & $\cdots$ & $\cdots$ & $\cdots$ \\
& M2 & $1893$ & $110$ & $ -0.04$ & $\cdots$ & $\cdots$ & $\cdots$ & $  0.63$ & $\cdots$ & $\cdots$ & $\cdots$ \\
& M3 & $1893$ & $106$ & $ -0.03$ & $\cdots$ & $\cdots$ & $\cdots$ & $  0.14$ & $\cdots$ & $\cdots$ & $\cdots$ \\
$N(1900)P_{13}$ & M0 & $1930$ & $158$ & $ -0.17$ & $  0.09$ & $ -0.13$ & $ -0.00$ & $  0.00$ & $  0.00$ & $  0.00$ & $  0.00$ \\
& M1 & $1930$ & $152$ & $ -0.16$ & $  0.08$ & $ -0.12$ & $  0.00$ & $  0.36$ & $ -0.24$ & $  0.30$ & $  2.00$ \\
& M2 & $1930$ & $161$ & $ -0.16$ & $  0.09$ & $ -0.12$ & $ -0.00$ & $ -1.80$ & $  0.71$ & $ -1.89$ & $  2.00$ \\
& M3 & $1930$ & $151$ & $ -0.14$ & $  0.08$ & $ -0.11$ & $ -0.00$ & $ -0.26$ & $ -0.51$ & $ -0.28$ & $ -2.00$ \\
$N(1990)F_{17}$ & M0 & $1995$ & $272$ & $-10.00$ & $  7.15$ & $ -8.16$ & $ -1.87$ & $  0.00$ & $  0.00$ & $  0.00$ & $  0.00$ \\
& M1 & $1995$ & $314$ & $-10.00$ & $  6.29$ & $ -7.98$ & $ -1.54$ & $  0.38$ & $  0.76$ & $  0.43$ & $  0.14$ \\
& M2 & $1995$ & $263$ & $-10.00$ & $  6.35$ & $ -8.09$ & $ -1.38$ & $  2.00$ & $  1.87$ & $ -0.22$ & $ -1.95$ \\
& M3 & $1995$ & $265$ & $-10.00$ & $  5.94$ & $ -8.03$ & $ -1.25$ & $  0.42$ & $  0.19$ & $  0.35$ & $ -1.96$ \\
$N(2000)F_{15}$ & M0 & $2090$ & $335$ & $ -1.29$ & $-10.00$ & $ -0.66$ & $  4.44$ & $  0.00$ & $  0.00$ & $  0.00$ & $  0.00$ \\
& M1 & $2090$ & $335$ & $ -1.26$ & $-10.00$ & $ -0.56$ & $  4.45$ & $ -0.85$ & $  0.51$ & $ -2.00$ & $  0.78$ \\
& M2 & $2090$ & $335$ & $ -1.12$ & $-10.00$ & $ -0.32$ & $  4.45$ & $ -2.00$ & $ -2.00$ & $  1.93$ & $  2.00$ \\
& M3 & $2090$ & $338$ & $ -1.28$ & $-10.00$ & $ -0.50$ & $  4.45$ & $ -0.68$ & $  1.94$ & $  0.38$ & $  1.80$ \\
$N(2060)D_{15}$ & M0 & $2060$ & $450$ & $  5.05$ & $  0.42$ & $  4.16$ & $ -1.39$ & $  0.00$ & $  0.00$ & $  0.00$ & $  0.00$ \\
& M1 & $2060$ & $450$ & $  5.03$ & $  0.48$ & $  4.14$ & $ -1.33$ & $  0.27$ & $  2.00$ & $  0.40$ & $  1.23$ \\
& M2 & $2060$ & $450$ & $  5.00$ & $  0.42$ & $  4.12$ & $ -1.40$ & $ -1.94$ & $ -1.99$ & $ -1.58$ & $ -2.00$ \\
& M3 & $2060$ & $450$ & $  4.95$ & $  0.43$ & $  4.08$ & $ -1.41$ & $ -0.93$ & $ -2.00$ & $ -0.99$ & $  0.38$ \\
$N(2120)D_{13}$ & M0 & $2075$ & $375$ & $ -0.08$ & $  1.70$ & $ -0.22$ & $ -0.67$ & $  0.00$ & $  0.00$ & $  0.00$ & $  0.00$ \\
& M1 & $2075$ & $375$ & $ -0.08$ & $  1.72$ & $ -0.22$ & $ -0.68$ & $ -0.22$ & $  0.23$ & $  0.05$ & $  0.22$ \\
& M2 & $2075$ & $375$ & $ -0.08$ & $  1.98$ & $ -0.21$ & $ -0.79$ & $  2.00$ & $  1.88$ & $  2.00$ & $  1.77$ \\
& M3 & $2075$ & $375$ & $ -0.08$ & $  1.86$ & $ -0.23$ & $ -0.74$ & $ -2.00$ & $  0.57$ & $ -0.20$ & $  0.59$ \\
$N(2190)G_{17}$ & M0 & $2175$ & $300$ & $ 10.00$ & $-10.00$ & $  3.02$ & $  6.49$ & $  0.00$ & $  0.00$ & $  0.00$ & $  0.00$ \\
& M1 & $2174$ & $300$ & $ 10.00$ & $-10.00$ & $  1.88$ & $  6.51$ & $  0.96$ & $  0.95$ & $  2.00$ & $  0.79$ \\
& M2 & $2181$ & $300$ & $ 10.00$ & $-10.00$ & $  2.54$ & $  6.45$ & $  2.00$ & $ -1.57$ & $  0.06$ & $  1.25$ \\
& M3 & $2181$ & $300$ & $ 10.00$ & $-10.00$ & $  1.78$ & $  6.34$ & $  0.58$ & $  2.00$ & $  0.87$ & $  1.55$ \\
$N(2220)H_{19}$ & M0 & $2200$ & $500$ & $-10.00$ & $-10.00$ & $ 10.00$ & $ -5.31$ & $  0.00$ & $  0.00$ & $  0.00$ & $  0.00$ \\
& M1 & $2200$ & $380$ & $-10.00$ & $-10.00$ & $ 10.00$ & $ -1.18$ & $  2.00$ & $  2.00$ & $  2.00$ & $  2.00$ \\
& M2 & $2200$ & $500$ & $-10.00$ & $-10.00$ & $ 10.00$ & $ -6.81$ & $ -1.33$ & $ -0.69$ & $  0.32$ & $  0.28$ \\
& M3 & $2200$ & $500$ & $-10.00$ & $-10.00$ & $ 10.00$ & $ -6.51$ & $  2.00$ & $  1.73$ & $  2.00$ & $  2.00$ \\
$N(2250)G_{19}$ & M0 & $2250$ & $300$ & $ 10.00$ & $ -9.60$ & $  6.08$ & $ 10.00$ & $  0.00$ & $  0.00$ & $  0.00$ & $  0.00$ \\
& M1 & $2250$ & $300$ & $ 10.00$ & $-10.00$ & $  5.32$ & $ 10.00$ & $ -1.12$ & $ -2.00$ & $ -2.00$ & $ -0.47$ \\
& M2 & $2253$ & $300$ & $ 10.00$ & $ -9.53$ & $  6.30$ & $ 10.00$ & $  1.66$ & $  1.30$ & $ -1.79$ & $ -0.48$ \\
& M3 & $2283$ & $300$ & $ 10.00$ & $ -6.58$ & $  6.25$ & $ 10.00$ & $  0.50$ & $ -0.69$ & $  1.41$ & $  0.94$ \\
    \end{tabular}
  \end{ruledtabular}
\end{table*}

In Table~\ref{tab:background} we list the leading coupling constants and
a number of background parameters obtained in the previous work (model M0
\cite{samson_thesis}) and the present analysis (models M1$-$M3).
Table~\ref{tab:background} unveils the difference between the effects
of CLAS and MAMI data. By comparing the parameters of M0 to those of M1
and M2 we can see that the changes are mainly in opposite directions. For instance,
including the CLAS (MAMI) data increases (decreases) the coupling constants
$g_{K \Lambda N}$, $G^{V}_{K_1}$, and the ratio $r_{K_1K\gamma}$. The opposite
effect is observed in the coupling constant $G^{T}_{K_1}$. These changes indicate
that substantial but different adjustments in the background sector are required 
to fit the $\gamma n\to K^0\Lambda$  CLAS and MAMI data. 
Furthermore, from the value of
$\chi^2/N_{\rm dof}$ it is seen that the MAMI data are slightly difficult to 
fit. As expected, including both data sets simultaneously results in moderate
coupling constants and other parameters. 

From Table~\ref{tab:background}
it is also important to note that both background and resonance cutoffs 
are almost unaffected by the inclusion of $\gamma n\to K^0\Lambda$ data, 
although significant suppression is required to bring the cross section 
in this channel to the right value. Thus, we might safely conclude that
this task is handled by the readjustment of coupling constants.

\begin{table}[!h]
 \centering
  \caption{Properties of the hyperon resonances extracted from different models.
      Note that for the sake of simplicity M refers to the specific model.}
  \label{tab:mass_hyp_extracted}
  \begin{ruledtabular}
    \begin{tabular}[c]{lcccrrrr}
      Resonance & M& $m_{Y^*}$& $\Gamma_{Y^*}$& $G^{(1)}_{Y^*}$ & $G^{(2)}_{Y^*}$ & $G^{(3)}_{Y^*}$ & $G^{(4)}_{Y^*}$ \\
      & &  (MeV) &  (MeV) & &&&\\
      \hline
$\Lambda(1405)S_{01}$ & M0 & $1404$ & $052$ & $-10.00$ & $\cdots$ & $\cdots$ & $\cdots$ \\
& M1 & $1404$ & $049$ & $ -8.99$ & $\cdots$ & $\cdots$ & $\cdots$ \\
& M2 & $1404$ & $052$ & $-10.00$ & $\cdots$ & $\cdots$ & $\cdots$ \\
& M3 & $1404$ & $052$ & $-10.00$ & $\cdots$ & $\cdots$ & $\cdots$ \\
$\Lambda(1520)D_{03}$ & M0 & $1518$ & $017$ & $  1.33$ & $ -9.29$ & $ -8.70$ & $  1.37$ \\
& M1 & $1518$ & $017$ & $  1.15$ & $ -9.19$ & $ -8.04$ & $  1.21$ \\
& M2 & $1518$ & $017$ & $  0.91$ & $ -9.32$ & $ -8.65$ & $  1.41$ \\
& M3 & $1518$ & $017$ & $  0.64$ & $ -9.39$ & $ -8.38$ & $  1.10$ \\
$\Lambda(1600)P_{01}$ & M0 & $1700$ & $250$ & $-10.00$ & $\cdots$ & $\cdots$ & $\cdots$ \\
& M1 & $1700$ & $250$ & $-10.00$ & $\cdots$ & $\cdots$ & $\cdots$ \\
& M2 & $1676$ & $250$ & $-10.00$ & $\cdots$ & $\cdots$ & $\cdots$ \\
& M3 & $1700$ & $250$ & $-10.00$ & $\cdots$ & $\cdots$ & $\cdots$ \\
$\Lambda(1670)S_{01}$ & M0 & $1680$ & $050$ & $  4.56$ & $\cdots$ & $\cdots$ & $\cdots$ \\
& M1 & $1660$ & $020$ & $ -1.76$ & $\cdots$ & $\cdots$ & $\cdots$ \\
& M2 & $1680$ & $020$ & $  7.43$ & $\cdots$ & $\cdots$ & $\cdots$ \\
& M3 & $1680$ & $020$ & $  5.65$ & $\cdots$ & $\cdots$ & $\cdots$ \\
$\Lambda(1690)D_{03}$ & M0 & $1685$ & $070$ & $ 10.00$ & $ 10.00$ & $  9.99$ & $  3.69$ \\
& M1 & $1685$ & $070$ & $  9.60$ & $ 10.00$ & $ 10.00$ & $  3.80$ \\
& M2 & $1685$ & $070$ & $  9.78$ & $ 10.00$ & $ 10.00$ & $  3.62$ \\
& M3 & $1685$ & $070$ & $ 10.00$ & $ 10.00$ & $  9.99$ & $  4.06$ \\
$\Lambda(1800)S_{01}$ & M0 & $1850$ & $400$ & $ 10.00$ & $\cdots$ & $\cdots$ & $\cdots$ \\
& M1 & $1720$ & $400$ & $ 10.00$ & $\cdots$ & $\cdots$ & $\cdots$ \\
& M2 & $1849$ & $400$ & $ 10.00$ & $\cdots$ & $\cdots$ & $\cdots$ \\
& M3 & $1850$ & $400$ & $  9.99$ & $\cdots$ & $\cdots$ & $\cdots$ \\
$\Lambda(1810)P_{01}$ & M0 & $1750$ & $050$ & $  6.84$ & $\cdots$ & $\cdots$ & $\cdots$ \\
& M1 & $1750$ & $050$ & $  4.94$ & $\cdots$ & $\cdots$ & $\cdots$ \\
& M2 & $1750$ & $050$ & $  1.63$ & $\cdots$ & $\cdots$ & $\cdots$ \\
& M3 & $1750$ & $050$ & $  2.43$ & $\cdots$ & $\cdots$ & $\cdots$ \\
$\Lambda(1890)P_{03}$ & M0 & $1850$ & $060$ & $ -3.78$ & $  6.89$ & $ 10.00$ & $  1.87$ \\
& M1 & $1850$ & $060$ & $ -2.43$ & $  7.86$ & $ 10.00$ & $  1.20$ \\
& M2 & $1850$ & $060$ & $ -3.99$ & $  8.37$ & $ 10.00$ & $  1.27$ \\
& M3 & $1850$ & $060$ & $ -3.52$ & $  9.91$ & $ 10.00$ & $  0.89$ \\
$\Sigma(1385)P_{13}$ & M0 & $1383$ & $031$ & $  0.57$ & $ -0.66$ & $ -1.02$ & $ -0.53$ \\
& M1 & $1383$ & $031$ & $  0.48$ & $ -0.72$ & $ -1.02$ & $ -0.48$ \\
& M2 & $1383$ & $031$ & $  0.60$ & $ -0.77$ & $ -1.02$ & $ -0.45$ \\
& M3 & $1383$ & $031$ & $  0.55$ & $ -0.88$ & $ -0.99$ & $ -0.42$ \\
$\Sigma(1660)P_{11}$ & M0 & $1630$ & $040$ & $ 10.00$ & $\cdots$ & $\cdots$ & $\cdots$ \\
& M1 & $1630$ & $040$ & $ 10.00$ & $\cdots$ & $\cdots$ & $\cdots$ \\
& M2 & $1630$ & $040$ & $  9.91$ & $\cdots$ & $\cdots$ & $\cdots$ \\
& M3 & $1630$ & $040$ & $ 10.00$ & $\cdots$ & $\cdots$ & $\cdots$ \\
$\Sigma(1670)D_{13}$ & M0 & $1685$ & $040$ & $-10.00$ & $ 10.00$ & $ 10.00$ & $ -6.89$ \\
& M1 & $1685$ & $040$ & $ -9.93$ & $ 10.00$ & $ 10.00$ & $ -6.68$ \\
& M2 & $1685$ & $040$ & $ -8.53$ & $ 10.00$ & $ 10.00$ & $ -6.91$ \\
& M3 & $1685$ & $040$ & $ -8.30$ & $ 10.00$ & $ 10.00$ & $ -6.45$ \\
    \end{tabular}
  \end{ruledtabular}
\end{table}

\begin{table}[!]
  \addtocounter{table}{-1}
 \centering
  \caption{Properties of the hyperon resonances extracted from different models (continued).}
  \begin{ruledtabular}
    \begin{tabular}[c]{lcccrrrr}
      Resonance & M& $m_{Y^*}$& $\Gamma_{Y^*}$& $G^{(1)}_{Y^*}$ & $G^{(2)}_{Y^*}$ & $G^{(3)}_{Y^*}$ & $G^{(4)}_{Y^*}$ \\
      & &  (MeV) &  (MeV) & &&&\\
      \hline
$\Sigma(1750)S_{11}$ & M0 & $1800$ & $160$ & $ 10.00$ & $\cdots$ & $\cdots$ & $\cdots$ \\
& M1 & $1730$ & $160$ & $ 10.00$ & $\cdots$ & $\cdots$ & $\cdots$ \\
& M2 & $1800$ & $060$ & $ 10.00$ & $\cdots$ & $\cdots$ & $\cdots$ \\
& M3 & $1800$ & $083$ & $ 10.00$ & $\cdots$ & $\cdots$ & $\cdots$ \\
$\Sigma(1880)P_{11}$ & M0 & $1804$ & $359$ & $-10.00$ & $\cdots$ & $\cdots$ & $\cdots$ \\
& M1 & $1804$ & $359$ & $-10.00$ & $\cdots$ & $\cdots$ & $\cdots$ \\
& M2 & $1804$ & $359$ & $-10.00$ & $\cdots$ & $\cdots$ & $\cdots$ \\
& M3 & $1804$ & $359$ & $-10.00$ & $\cdots$ & $\cdots$ & $\cdots$ \\
$\Sigma(1940)D_{13}$ & M0 & $1900$ & $300$ & $-10.00$ & $ 10.00$ & $  2.76$ & $  0.99$ \\
& M1 & $1950$ & $300$ & $-10.00$ & $ 10.00$ & $ -1.26$ & $  1.41$ \\
& M2 & $1903$ & $254$ & $ -9.99$ & $ 10.00$ & $  0.70$ & $  0.77$ \\
& M3 & $1917$ & $287$ & $-10.00$ & $ 10.00$ & $ -1.13$ & $  0.33$ \\
$\Sigma(2080)P_{13}$ & M0 & $2084$ & $234$ & $-10.00$ & $-10.00$ & $ 10.00$ & $ 10.00$ \\
& M1 & $2084$ & $234$ & $-10.00$ & $-10.00$ & $  7.17$ & $ 10.00$ \\
& M2 & $2084$ & $234$ & $-10.00$ & $-10.00$ & $  9.93$ & $ 10.00$ \\
& M3 & $2084$ & $234$ & $-10.00$ & $-10.00$ & $  8.75$ & $ 10.00$ \\
    \end{tabular}
  \end{ruledtabular}
\end{table}

The extracted nucleon resonance properties for all four models are given in 
Table~\ref{tab:mass_nuc_extracted}. Note that during the fitting process
the mass and width of resonances are constrained within the corresponding uncertainties 
given by PDG \cite{pdg}. This constraint is especially important for the present single 
channel analysis, since during the fitting process the extracted mass and width of 
resonances could become unrealistic. Such a problem does not appear in the
coupled-channels analysis, in which other channels work simultaneously to constrain
the mass and width of resonances. Since the values listed by PDG are mostly obtained
from coupled-channels analyses, by using the PDG values as the constraint we believe that 
we have partly incorporated this concern in our present work.

We notice that there are no dramatic changes in the masses and width of the resonances
after the inclusion of the $K^0\Lambda$ data. Only the mass of $N(1875)D_{13}$ resonance
changes rather significantly. The same situation also happens in the case of coupling
constants $G_{N^*}^{(i)}$. Therefore, the difference between $K^+\Lambda$ and $K^0\Lambda$
observables is mainly controlled by the ratio $r_{N^*}^{(i)}$ defined by
Eq.~(\ref{eq:nucl-res-trans-moment1}). Furthermore, the large coupling constants shown
in Table~\ref{tab:mass_nuc_extracted} do not directly indicate the importance of these
resonances. We will discuss the latter in the next subsection.

\begin{figure*}[!]
  \includegraphics[scale=0.7]{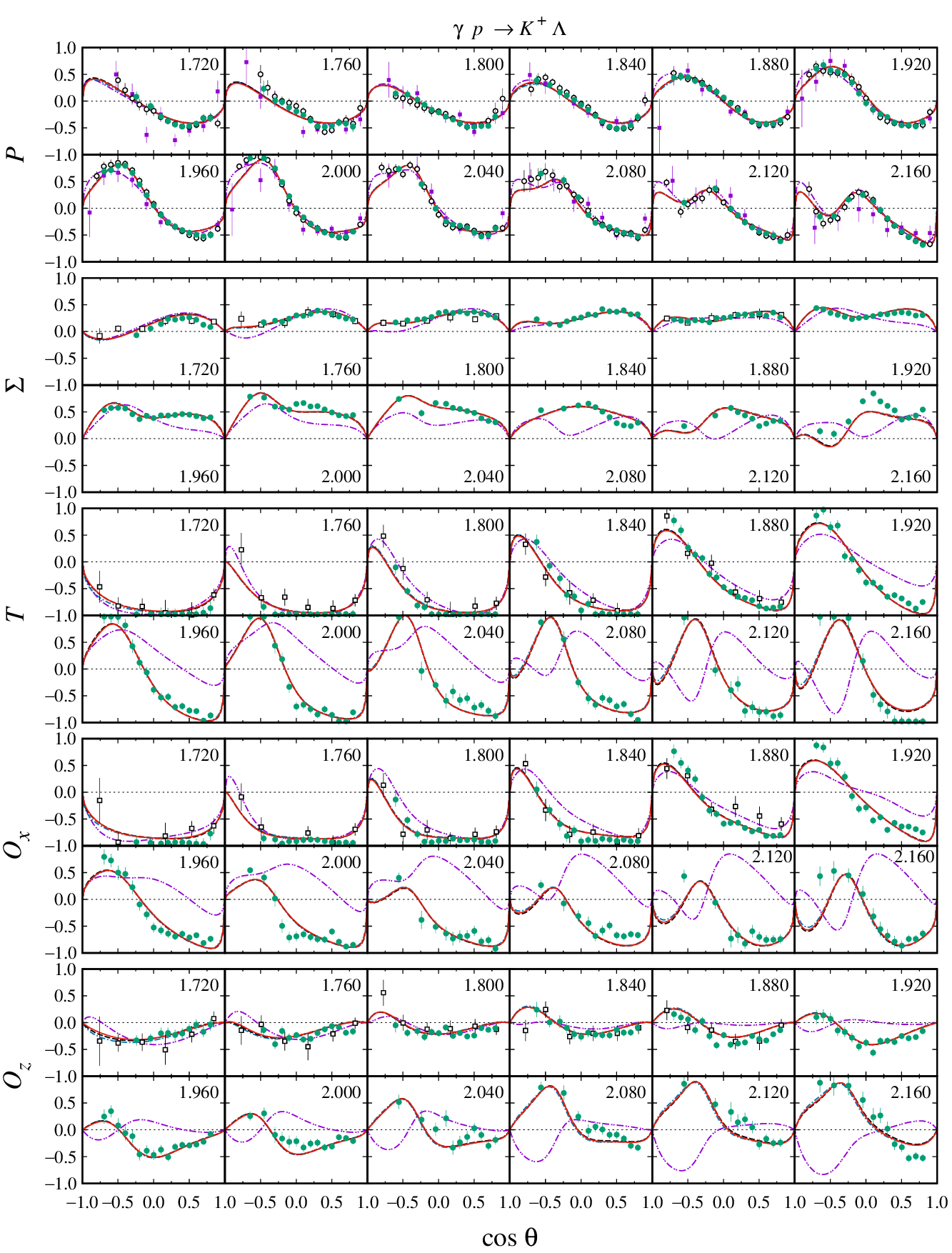}
\caption{Angular distributions of the recoil polarization $P$, 
  photon asymmetry $\Sigma$, target asymmetry $T$, and 
  photon-recoil double polarizations $O_x$ and $O_z$ for the
  $\gamma p \to K^+\Lambda$ process obtained
  from the previous multipole model (dot-dashed curves
  \cite{Mart:2017mwj}), the original field theoretic model
  involving the nucleon resonances with spins up to 9/2 (dashed 
  curves \cite{Clymton:2017nvp}), and the present work (solid
  curves). The experimental data shown in this figure are taken from 
  the GRAAL 2007 (open squares \cite{lleres:2007}), 
  CLAS 2006 (solid squares \cite{Bradford:2005pt}), 
  CLAS 2010 (open circles \cite{mcCracken}), and 
  CLAS 2016 (solid circles \cite{paterson}) collaborations.}
\label{fig:clas_new_pol}
\end{figure*}

In Table~\ref{tab:mass_hyp_extracted} we list the extracted masses, widths, and coupling
constants of the hyperon resonances, which contribute to the background terms. The same
phenomena as in the nucleon resonances are also seen in this case; i.e., there are 
almost no dramatic changes in these parameters except in the case of the $\Sigma(1940)D_{13}$ 
state. Presumably, these changes are required to explain the polarization observables
shown in Fig.~\ref{fig:clas_new_pol}. We also comeback to this topic later.

\begin{figure*}[!]
  \includegraphics[scale=0.70]{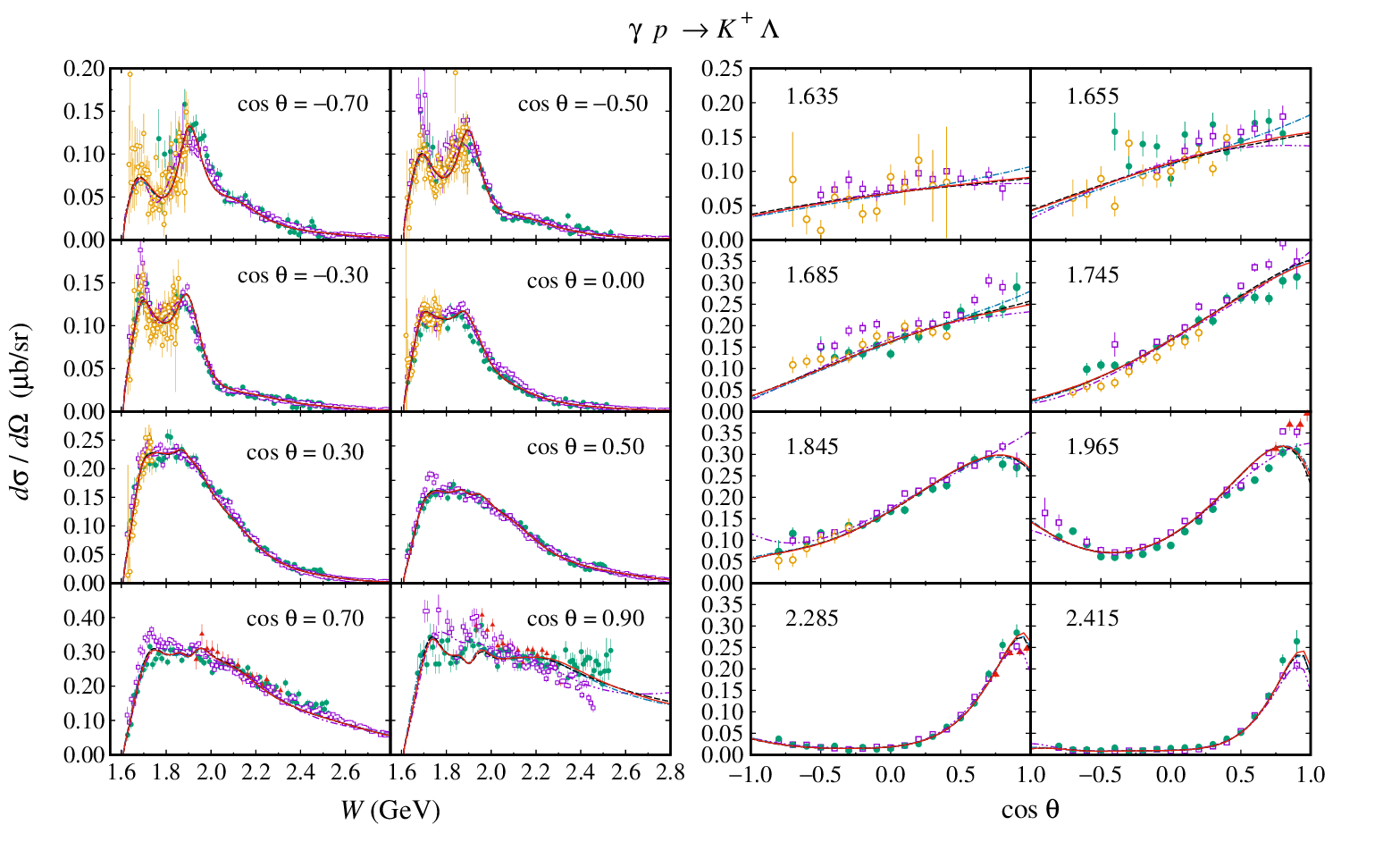}
\caption{Energy and angular distributions of the 
  $\gamma p \to K^+\Lambda$ differential cross section
  obtained in the present work and from the latest experimental
  data. Data shown in this figure are taken 
  from the LEPS 2006 (solid triangles \cite{Sumihama06}), 
  CLAS 2006 (solid squares \cite{Bradford:2005pt}), 
  CLAS 2010 (solid circles \cite{mcCracken}), and 
  Crystal Ball 2014 (open circles \cite{Jude:2013jzs})
  collaborations. For notation of the curves see 
  Fig.~\ref{fig:clas_new_pol}. }
\label{fig:dkple}
\end{figure*}

Figure \ref{fig:dkple} displays the energy and angular distributions
of the  $\gamma p\to K^+\Lambda$ differential cross section obtained from 
all four models (M0, M1, M2, and M3) and compared with presently 
available experimental data. 
As expected the difference between these models is almost negligible, except
in very forward direction. The origin of this difference is also 
obvious, i.e., experimental data in forward direction are more scattered 
than in other directions (see the panel for $\cos\theta=0.90$ in 
Fig.~\ref{fig:dkple}). Therefore, in this kinematics constraint from 
experimental data is less stringent during the fitting process and, 
as a result, the variance between the models becomes more apparent. 
We note that the same phenomenon, but only in a certain energy region, 
is also observed in the backward direction (see the panel for 
$W=1.845$ GeV in Fig.~\ref{fig:dkple}).

For the recoil polarization $P$ shown in Fig.~\ref{fig:clas_new_pol}
all models can nicely reproduce the new and older data. This is not surprising
because the recoil polarization data have been available since the last
decades, whereas the new CLAS 2016 data \cite{paterson} are consistent with
the older ones. The same result is also shown by the photon  $\Sigma$ 
and target $T$ asymmetries,
where we can see that our previous multipole model  \cite{Mart:2017mwj} 
(shown by the dot-dashed curves in Fig.~\ref{fig:clas_new_pol}) can easily
fit the low energy data but fails to reproduce the higher energy ones, since
it was fitted to the the GRAAL 2007 data \cite{lleres:2007} which are available
for $W$ only up to 1.9 GeV. A similar situation also occurs in the
case of double polarizations $O_x$ and $O_z$.

\begin{figure}
  \includegraphics[scale=0.372]{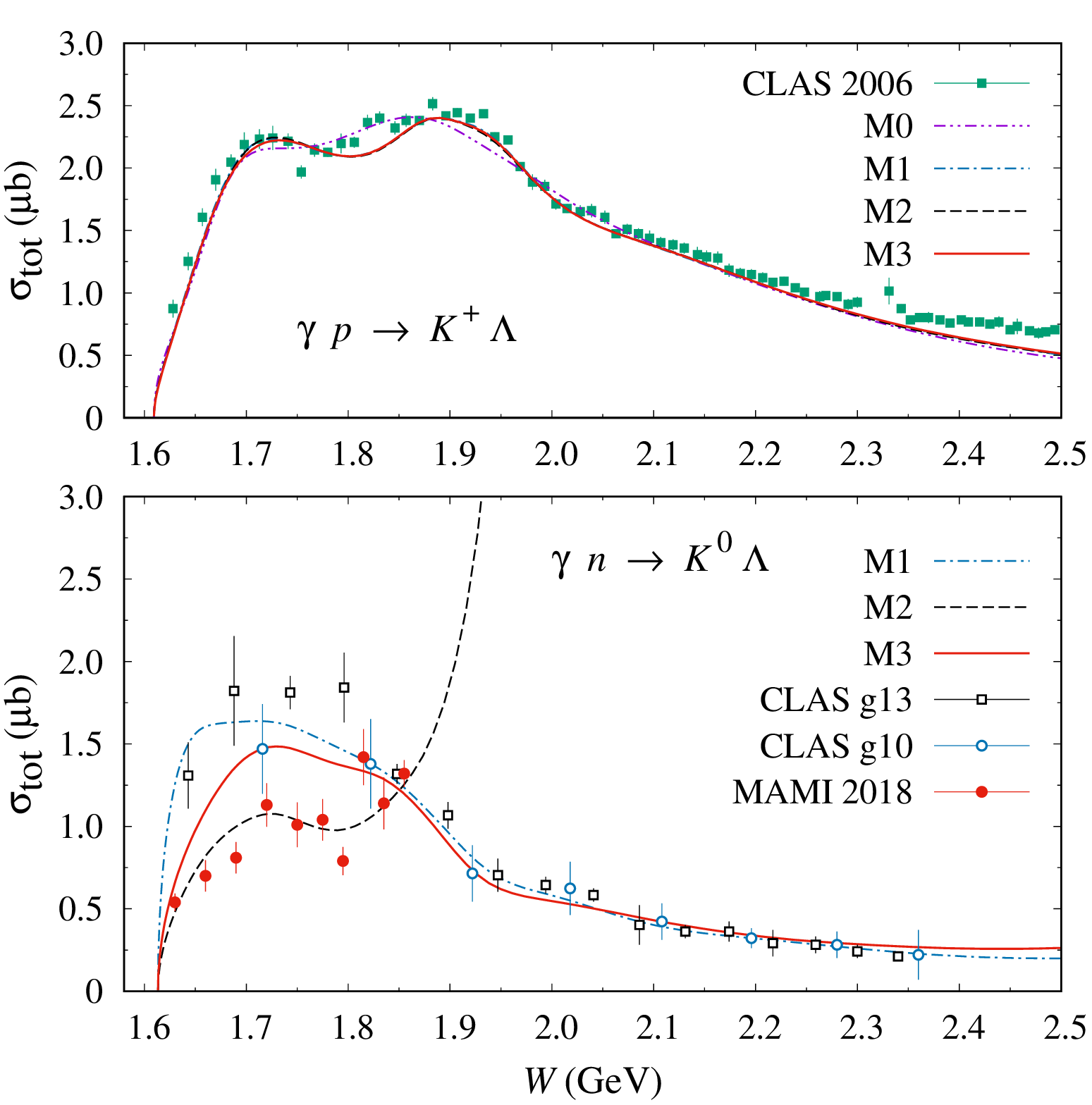}
\caption{As in Fig.~\ref{fig:tot_prob} but for the previous  \cite{Clymton:2017nvp}
  (M0) and present (M1$-$M3) analyses.
  Note that the previous analysis \cite{Clymton:2017nvp} did not predict
  the $\gamma n \to K^0\Lambda$ total cross section.}
\label{fig:k0total}
\end{figure}

In summary, our present models can nicely reproduce experimental data for 
the $\gamma p\to K^+\Lambda$ channel. This is expected because the difference
between the models originates from the $\gamma n\to K^0\Lambda$ channel.
Pictorially, the former is illustrated in the upper panel of Fig.~\ref{fig:k0total},
where we can see that only the model M0 slightly differs from the other models.
We have discussed the reason behind this phenomenon; i.e., the corresponding differential
cross section is slightly larger for $W< 1.9$ GeV but turns out to be smaller as
$W\gtrsim 1.9$ GeV. 

The situation is very different in the case of the $\gamma n\to K^0\Lambda$ channel.
As shown in the bottom panel of Fig.~\ref{fig:k0total} the inclusion of MAMI data
(model M2) results in a divergent total cross section for $W\gtrsim 1.9$ GeV, in 
contrast to the other models. Obviously this result is caused by the MAMI data, 
which are limited only up to 1.855 GeV. Above this energy region there is 
practically no constraint for the $\gamma n\to K^0\Lambda$ cross section.
By adding the CLAS data to this result we obtain model M3 which nicely 
reproduces all data for $W>1.8$ GeV and yields a compromise total cross
section for $W<1.8$ GeV. 

\begin{figure*}
  \includegraphics[scale=0.70]{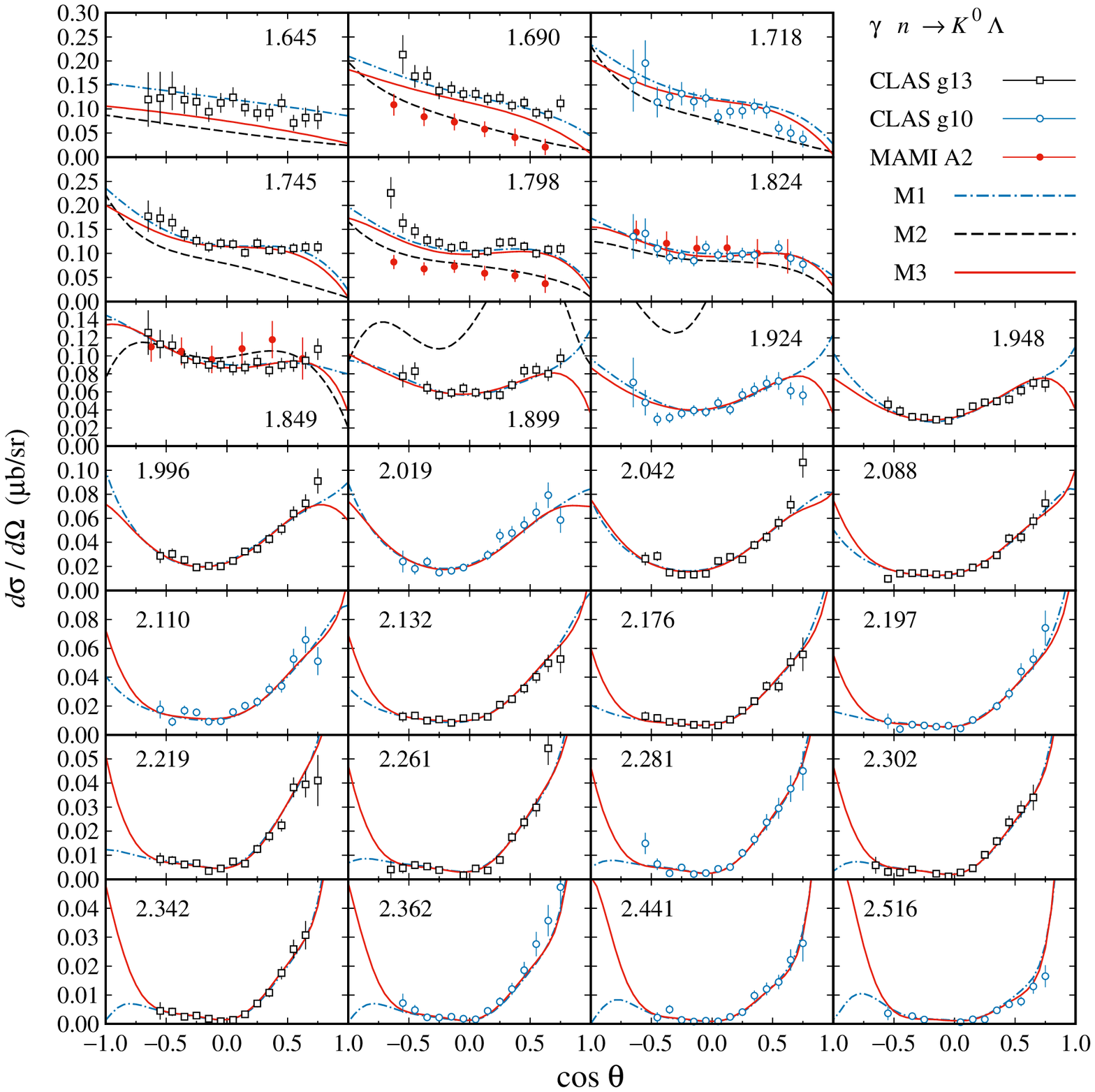}
\caption{As in the bottom panel of Fig.~\ref{fig:k0total}, but for the
         angular distribution of $\gamma n\to K^0\Lambda$ 
         differential cross section.}
\label{fig:k0lam_th}
\end{figure*}

\begin{figure*}
  \includegraphics[scale=0.70]{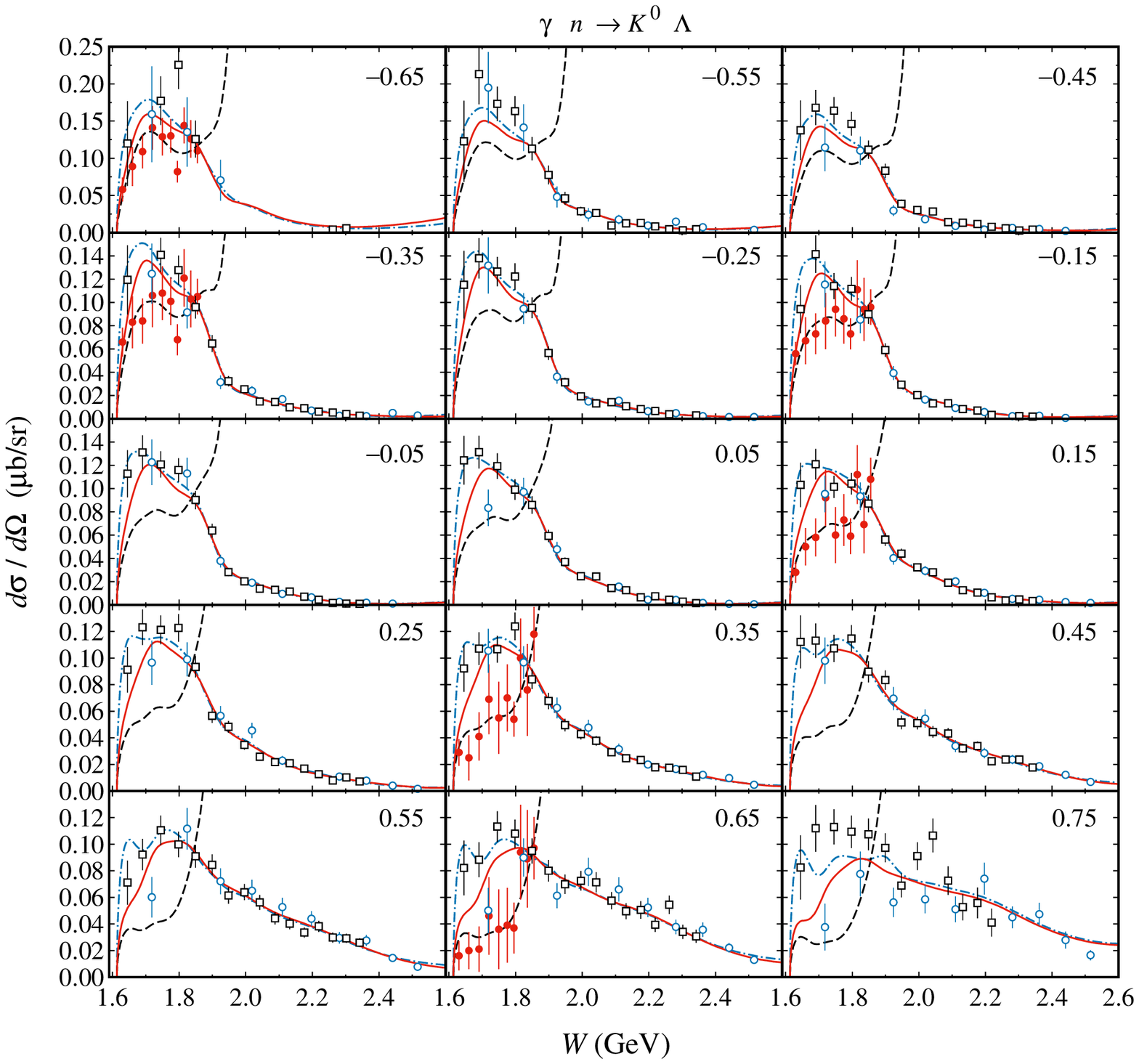}
\caption{As in Fig.~\ref{fig:k0lam_th}, but for the
         energy distribution.}
\label{fig:k0lam_w}
\end{figure*}

The angular and energy distributions of the differential cross section shown in
Figs. \ref{fig:k0lam_th} and \ref{fig:k0lam_w}, respectively, reveal more
information. Whereas the difference between models M1 and M2 is clear, the 
difference between models M1 and M3 is observed only in the backward region and,
for certain energies, in the forward region (see Fig. \ref{fig:k0lam_th}). 
It is well known that this
behavior originates from the $t$- and $u$-channel contributions that are
different for models M1 and M3 as we can see from the corresponding 
values of $r_{K_1K\gamma}$ in Table~\ref{tab:background} and $G^{(i)}_{Y^*}$ 
of Table~\ref{tab:mass_hyp_extracted}. From the energy distribution of the
differential cross section shown in Fig. \ref{fig:k0lam_w} we can understand
that the problem of data discrepancy is very clear in the forward direction,
where even the CLAS g10 and g13 data show variance for $\cos\theta=0.75$. 
In the future we expect that experimental measurement of kaon photoproduction 
should focus on the forward region in order to reconcile this discrepancy 
as well as to resolve the same problem in the case of the $K^+\Lambda$
photoproduction (see, e.g., Ref.~\cite{Mart:2017mwj}).

\subsection{Beam-target helicity asymmetry $E$ in the $\gamma n\to K^0\Lambda$ photoproduction}

\begin{figure*}
  \includegraphics[scale=0.40]{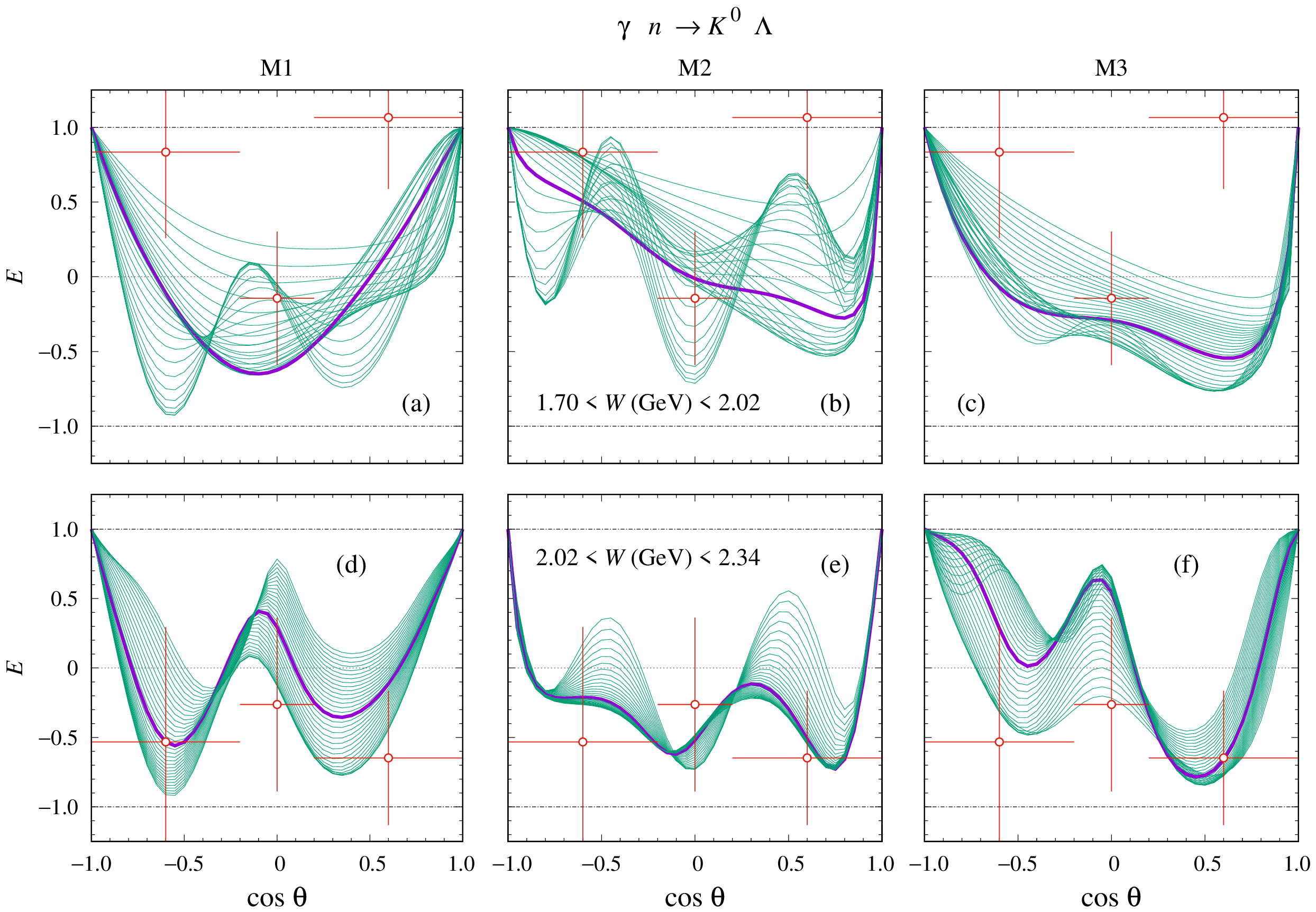}
\caption{Angular distributions of the $\gamma n\to K^0\Lambda$ helicity asymmetry $E$ for two
         different energy bins. Experimental data are taken from Ref.~\cite{Ho:2018riy}. 
         Thin solid curves are predictions of models M1$-$M3 calculated with 10 MeV 
         energy step and $W$ from 1.70 to 2.02 GeV (from 2.02 to 2.34 GeV) in the panels (a)$-$(c)
         [(d)$-$(f)]. Thick solid curves are obtained by using the middle values of
         the energy bins.}
\label{fig:pole_hd_th}
\end{figure*}

Recently, the CLAS collaboration~\cite{Ho:2018riy} has measured the beam-target helicity 
asymmetry $E$ in the $\gamma n\to K^0\Lambda$ process by using the CEBAF Large Acceptance 
Spectrometer on a 5-cm-long solid hydrogen deuteride target with the c.m. energy $W$ from
1.70 to 2.34 GeV. Due to the small cross section of the $K^0\Lambda$ final state and to detector 
inefficiencies the corresponding angular and energy bins are very large. The result is 
given in two energy bins, i.e., from 1.70 to 2.02 GeV and from 2.02 to 2.34 GeV. As shown in
Fig.~\ref{fig:pole_hd_th} each energy bin contains three data points. Nevertheless, since 
data on $K^0\Lambda$ photoproduction are very limited, these six data points are still invaluable
for our present work. However, due to their large error bars we did not include these data in the 
fitting database. Furthermore, during the fitting process these six data points could not compete 
with other data that have much smaller error bars, unless a weighting factor was introduced. 
Therefore, in the present work we only compare the predictions of the three proposed models
with these data. For the first energy bin we have calculated the asymmetry $E$ from 1.70 to 2.02 
GeV with energy step 10 MeV and compared the 33 calculated asymmetries with experimental data 
in panels a, b, and c of Fig.~\ref{fig:pole_hd_th}. We have repeated the calculation for the 
second energy bin and compared the result with data in panels d, e, and f of Fig.~\ref{fig:pole_hd_th}.
From all panels of Fig.~\ref{fig:pole_hd_th} we may conclude that the present work 
provides the asymmetry bands that are comparable with the experimental data for both energy bins. 

By considering the existing error bars shown in Fig.~\ref{fig:pole_hd_th} we might conclude 
that the six available data points are still compatible with all models. However, 
although their uncertainties are relatively large, these data still exhibit a clear trend. 
For the low energy bin (from 1.70 to 2.02 GeV) they show a minimum at $\cos\theta=0$. 
This is in contrast to the high energy bin (from 2.02 to 2.34 GeV), for which the 
asymmetry is maximum at $\cos\theta=0$. 
Thus, by comparing the results obtained from models M1 [panels (a) and (d)] and M2 
[panels (b) and (e)] we can see that model M1 is more consistent with the data. Panel (e)
shows that model M2 predicts a minimum helicity asymmetry near  $\cos\theta=0$, in contrast
to the experimental data. Perhaps this is not surprising, because model M1 was fitted
to the CLAS $\gamma n\to K^0\Lambda$ differential cross section, whereas model M2 was
fitted to the MAMI data. As expected, the use of both data sets yields a compromise
model M3. However, from panels (c) and (f) we can see that model M3 has s relatively similar 
trend to model M1. We believe that this occurs because in our fitting database the number of CLAS
$\gamma n\to K^0\Lambda$ differential cross section data is much larger than the MAMI ones
and as a consequence the latter have smaller influence.

To conclude this subsection we may safely say that the models that fit the CLAS 
$\gamma n\to K^0\Lambda$ data are more consistent with the presently available 
helicity asymmetry $E$ data measured by the CLAS collaboration. Nevertheless, 
more accurate experimental data are strongly required to support this conclusion. 

\subsection{The significance of individual resonances}
As in the previous analyses \cite{Mart:2006dk,Mart:2017mwj} in the present work
we also investigate the significance of each resonance used in the models by defining 
the parameter
\begin{eqnarray}
  \label{eq:res_contrib}
  \Delta \chi^2 ~=~ \frac{\chi^2_{{\rm All}-N^*}-\chi^2_{\rm All}}{\chi^2_{\rm All}}
  \times 100\,\% ~,
\end{eqnarray}
where $\chi^2_{\rm All}$ is the $\chi^2$ obtained if all nucleon resonances 
were used and $\chi^2_{{\rm All}-N^*}$ is the $\chi^2$ obtained if 
a specific nucleon resonance was excluded. Therefore, Eq.~(\ref{eq:res_contrib}) 
does not express the contribution of the specific resonance in the process, 
but it solely quantifies how difficult to reproduce experimental data without
this resonance. This is the reason why we call it the significance of the
resonance. As a further note, in our experience, although 
Eq.~(\ref{eq:res_contrib}) seems to be very simple, 
the numerical calculations to obtain the $\Delta \chi^2$ for each resonance
in all models requires high CPU time.

\begin{figure}[t]
\centering
  \centering\includegraphics[scale=0.50,angle=-90]{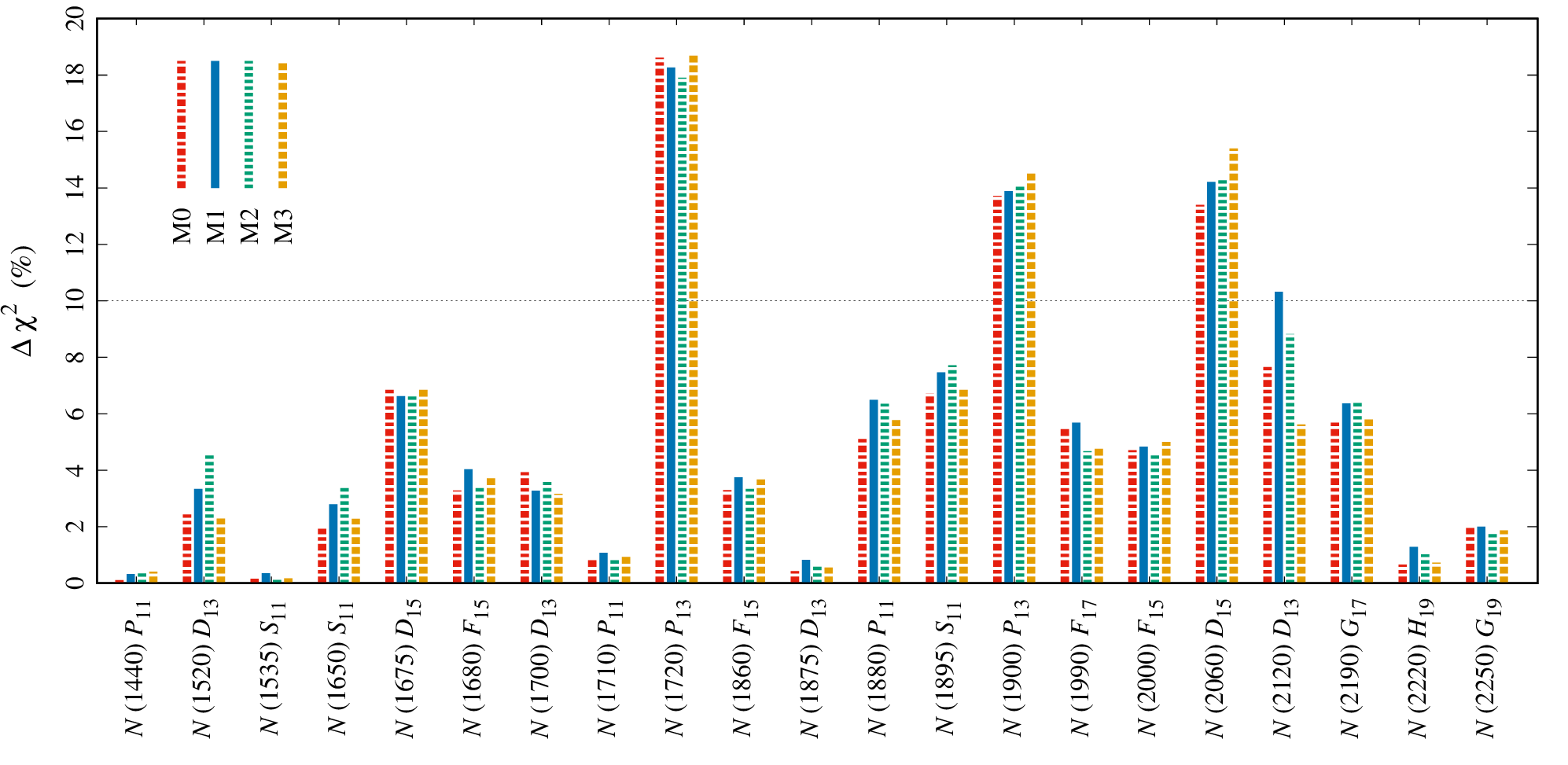}
 \caption{The significance of individual resonances calculated from 
  Eq.~(\ref{eq:res_contrib}) for all models investigated in the present
  work.}\label{fig:strength}
\end{figure}   

The result of previous investigation by using a multipole approach 
is relatively unstable to the choice of 
data sets included in the fitting database \cite{Mart:2006dk}.
For instance, the $N(1650)S_{11}$ resonance was found to be very important if we used
the SAPHIR data \cite{Tran:1998,Glander:2003jw}, but it turned out to be unimportant if 
the SAPHIR data were replaced with the CLAS data \cite{Bradford:2005pt}. The
latter would not change if both SAPHIR and CLAS data were used in the fitting 
process \cite{Mart:2006dk}. Nevertheless, in this multipole analysis 
a number of nucleon resonances were found to be important and relatively
stable to the choice of data sets. Included in this category are the 
$N(1720)P_{13}$, $N(1900)P_{13}$, and $N(2080)D_{13}$ resonances. 
Note that in the recent PDG listing the $N(2080)D_{13}$ state 
does not exist any longer, it has been replaced by the $N(1875)D_{13}$ 
state \cite{pdg}. Therefore, comparing the result for this resonance
to the present work is not possible. Moreover, the number of experimental
data used in this multipole approach is less than 2500, i.e., much smaller
than that used in the present work. 

In a more recent multipole analysis, by using the same experimental data points 
as in model M0 of the present work, it was found that the most significant nucleon 
resonances in the $K^+\Lambda$ photoproduction are the $N(1650)S_{11}$, 
$N(1720)P_{13}$, $N(1900)P_{13}$, and $N(2060)D_{15}$ states \cite{Mart:2017mwj}.
Thus, we believe that the result of this analysis is more conclusive than that
of the previous one \cite{Mart:2006dk}, especially because the resonance properties
were constrained within the uncertainties of PDG estimate as in the present work.

The result of our present work is depicted in Fig.~\ref{fig:strength}, where 
the significance of each resonance is calculated for all four models. 
Obviously, our present calculation is stable to the choice of the $K^0\Lambda$ 
data sets, in contrast to the result of previous study \cite{Mart:2006dk}. 
From Fig.~\ref{fig:strength} we might say that the three most important resonances 
are the $N(1720)P_{13}$, $N(1900)P_{13}$, and $N(2060)D_{15}$ ones. Thus, our present 
calculation corroborates the finding of our recent multipole analysis
\cite{Mart:2017mwj}, except for the $N(1650)S_{11}$ state. We note that, although
the $N(1650)S_{11}$ resonance is insignificant in all models, this resonance still
gives sizable contribution to both $K^+\Lambda$ and $K^0\Lambda$ channels. This can be
seen from either the $\Delta \chi^2$ shown in Fig.~\ref{fig:strength} or the coupling
constant $G^{(1)}_{N(1650)}$ given in Table \ref {tab:mass_nuc_extracted}.

The well-known $N(1710)P_{11}$ resonance is found to be insignificant in the
present work. This finding corroborates the result of previous works 
\cite{Mart:2006dk,Mart:2017mwj}. This result is however different from the 
PDG estimate that rates this resonance with four-star status 
and branching ratio $\Gamma(\Lambda K)/\Gamma_{\rm total}$ of up to 25\%
\cite{pdg}. However, in the present work this is not bad news since 
with the absence of the important $P_{11}$ state near the threshold we could 
expect the increase of the probability to find the $P_{11}$ 
narrow resonance in this energy region. We discuss this topic
in the following subsection.

\subsection{Narrow resonance in the $K^0\Lambda$ photoproduction}

The existence of the $J^p=1/2^+$ ($P_{11}$) narrow resonance has drawn much attention 
from the hadronic physics communities since it was predicted by the chiral quark soliton 
model as one of the ten members of the antidecuplet baryons \cite{diakonov}. These baryons 
are very interesting because three of them are exotic; i.e., their quantum numbers can 
only be constructed from five quarks or a pentaquark. The resonance was originally assigned to
the $N(1710)P_{11}$ state with the estimated width $\Gamma=41$ MeV, 
since information from PDG was uncertain \cite{pdg_old}.
However, after the reports of experimental observation of the exotic baryons $\Xi_{3/2}$ 
\cite{NA49} and $\Theta^+$ \cite{nakano}, Ref.~\cite{Walliser:2003dy} found that the 
$P_{11}$ mass should be either 1650 or 1660 MeV, depending on whether the symmetry 
breaker $\Delta$ was included or not, respectively. On the other hand, by using the masses 
of the two exotic baryons as inputs, the authors of Ref.~\cite{diakonov} reevaluated their 
prediction and found that the $P_{11}$ mass became either 1690 or 1647 MeV, if the mixing 
with the lower-lying nucleonlike octet was considered or not, respectively \cite{diakonov2004}. 

The chiral quark soliton model also predicted a large $\eta N$ branching ratio. As a result,
there was considerable interest in reevaluation of the $\eta$ photoproduction at energies 
around 1700 MeV. It was then reported that a substantial enhancement in the cross section of 
$\eta$ photoproduction off a free neutron has been observed at $W\approx 1670$ MeV 
\cite{kuznetsov}. Two experiments performed later by other collaborations confirmed  this finding
\cite{confirmed}. Interestingly, this enhancement is absent or very weak in the case of 
photoproduction off a proton target.

Another mechanism that can be used to study this narrow resonance is the $\pi N$ scattering 
and photoproduction, since the chiral quark soliton model also predicted a sizable decay width
to this channel. By using a modified partial wave analysis Ref.~\cite{igor}  obtained 
the $P_{11}$ mass from $\pi N$ data. This was achieved by scanning the changes of
$\chi^2$ (called $\Delta\chi^2$) in the range of resonance mass between 1620 and 1760 MeV 
after including this resonance in the $P_{11}$ partial wave. A relatively large $\Delta\chi^2$ 
was observed at 1680 MeV and a smaller one was found at 1730 MeV. This result was found to be 
independent of the total width and branching ratio of the resonance.

Motivated by the fact that both $N^*\to K\Lambda$ and  $N^*\to \pi N$ branching ratios 
are predicted by the chiral quark soliton model  \cite{diakonov}, we have investigated 
the existence of the narrow resonance in the $\gamma p \to K^+\Lambda$ channel by 
utilizing two isobar models which are able to describe the experimental data from 
threshold up to $W=1730$ MeV \cite{mart-narrow}. By analyzing the changes in the 
total $\chi^2$ with the variation of resonance mass from 1620 to 1730 MeV and resonance 
width from 0.1 to 1 MeV and from 1 to 10 MeV we found that the most promising candidate 
mass and width of this resonance are 1650 and 5 MeV, respectively \cite{mart-narrow}. 
However, there was
very small signal found in the total cross section since the effect of this resonance 
on differential cross section switches from decreasing to increasing as the kaon angle
increases. The net result in the total cross section is nearly 0. Interestingly,
it was found that the narrow resonance signal originates mostly from the $\Lambda$ 
recoil polarization data. As a consequence, further measurement of recoil polarization
with much smaller error bars is strongly recommended. 

Given the fact that the effect of this resonance on the cross section of 
$\eta$ photoproduction is more substantial in the neutron channel \cite{kuznetsov}, 
instead of the proton one, it is obviously important to investigate the effect on
the neutron channel of kaon photoproduction, i.e., the $\gamma n\to K^0\Lambda$
process. In the previous work \cite{mart-narrow} we fitted the $\gamma p\to K^+\Lambda$ 
experimental data to obtain the $\chi^2$ with fixed resonance mass and width and repeated 
the fitting process for different mass and width values. For the sake of simplicity, in
the present work we consider both resonance mass and width as free parameters.
Furthermore, since the resonance contribution to the scattering amplitude in the neutron 
channel is determined by the product of $g_{\gamma nN^*}\,  g_{K^{0} \Lambda N^*}$, 
we should also consider this product as a free parameter. However, 
Eq.~(\ref{eq:nucl-res-trans-moment1}) immediately tell us that this product is related 
to the $g_{\gamma pN^*}\,  g_{K^{+} \Lambda N^*}$ 
by the ratio $r_{N^*}$ because $g_{K^{+} \Lambda N^*} = g_{K^{0} \Lambda N^*}$ from
Eq.~(\ref{eq:lambda_coupling}). Therefore, for our purpose it is sufficient to extract 
the ratio $r_{N^*}$ from the fitting process.

\begin{table}[!]
  \centering
  \caption{Properties of the $P_{11}$ narrow resonance extracted in the present work 
    obtained from three different models.}
  \label{tab:narrow}
  \begin{ruledtabular}
    \centering
    \begin{tabular}{lrrr} 
      Parameter & M1 & M2 & M3 \\[1ex]
      \hline\\[-2.2ex]
      $m_{N^*}$ (MeV)      & $   1625    $ & $  1670   $ & $   1648 $ \\    
      $\Gamma_{N^*}$ (MeV) & $      2    $ & $    20   $ & $      3 $ \\    
      $r_{N^*}$      & $  -10.000  $ & $ -1.997  $ & $  6.696 $ \\[1ex]    
    \end{tabular}
  \end{ruledtabular}
\end{table}

The extracted ratios, masses, and widths of the $P_{11}$ narrow resonance obtained from 
the three different models M1$-$M3 are listed in Table~\ref{tab:narrow}.
Table~\ref{tab:narrow} reveals the substantially different results obtained from M1 and
M2 models, i.e., including the CLAS 2016 (\cite{paterson}) or MAMI 2018 \cite{Akondi:2018shh} 
data, respectively. Including the CLAS 2016 data yields the smallest ratio. Indeed, the 
extracted ratio reaches its lower bound. Since the cross section is proportional to the
squared amplitude, it is obvious that the model M1 yields a strong signal of narrow
resonance in the $\gamma n\to K^0\Lambda$ channel. The use of MAMI 2018 data yields a 
completely different result, where from Table~\ref{tab:narrow} we might expect a much
weaker signal in this case. However, the extracted mass and width are the largest for this
case. We note that if the MAMI data were used the extracted mass corroborated the
finding in $\eta$ photoproduction off a free neutron \cite{kuznetsov}, i.e., 1670 MeV.

The inclusion of both CLAS and MAMI data (model M3) yields a compromise result, where 
all extracted parameters fall between those of models M1 and M2, up to the sign of 
$r_{N^*}$. The effect of a different sign of $r_{N^*}$ in model M3 is discussed later.
Thus, the narrow resonance signal in the cross section is still sufficiently large.
We also note that the extracted mass and width in this case are closer to the finding 
in our previous work which only utilized the $\gamma p\to K^+\Lambda$ data, i.e.,
1650 and 5 MeV, respectively \cite{mart-narrow}.

\begin{figure*}
  \includegraphics[scale=0.40]{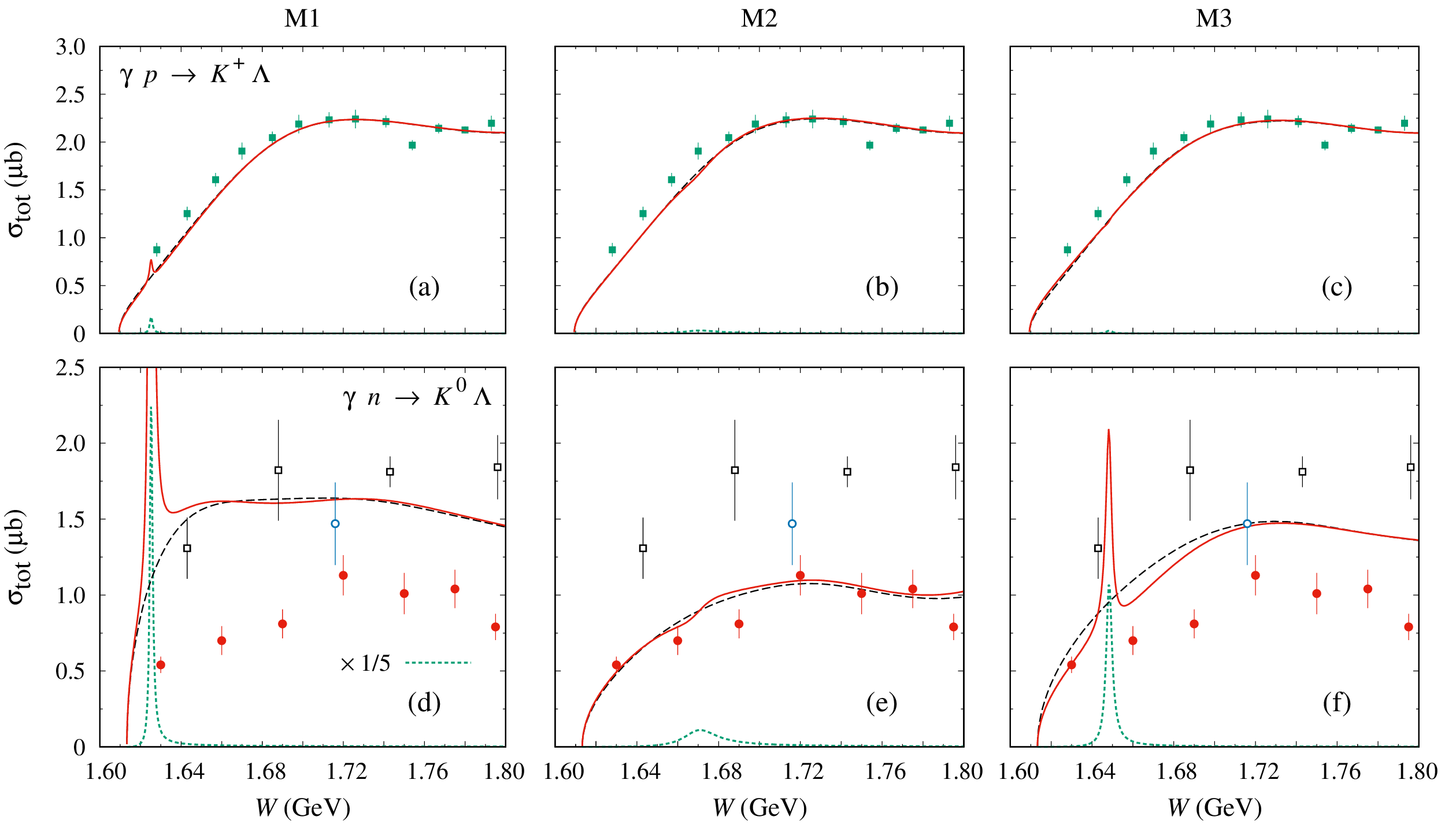}
\caption{Effects of the $P_{11}$ narrow nucleon resonance on the total cross section of
  the $\gamma p\to K^+\Lambda$ [panels (a)$-$(c)] and $\gamma n\to K^0\Lambda$ [panels (d)$-$(f)] 
  channels, obtained from three different models M$-$M3. Solid and dashed curves show the results
  obtained by including and excluding this resonance in these models, respectively.
  The dotted curves display contribution of this resonance alone. Experimental data are as in
  Fig.~\ref{fig:k0total}. Note that contribution of the $P_{11}$ narrow resonance in panel d 
  has been rescaled by a factor of 1/5 in order to fit in the same plot.}
\label{fig:narrow}
\end{figure*}

\begin{figure*}
  \includegraphics[scale=0.70]{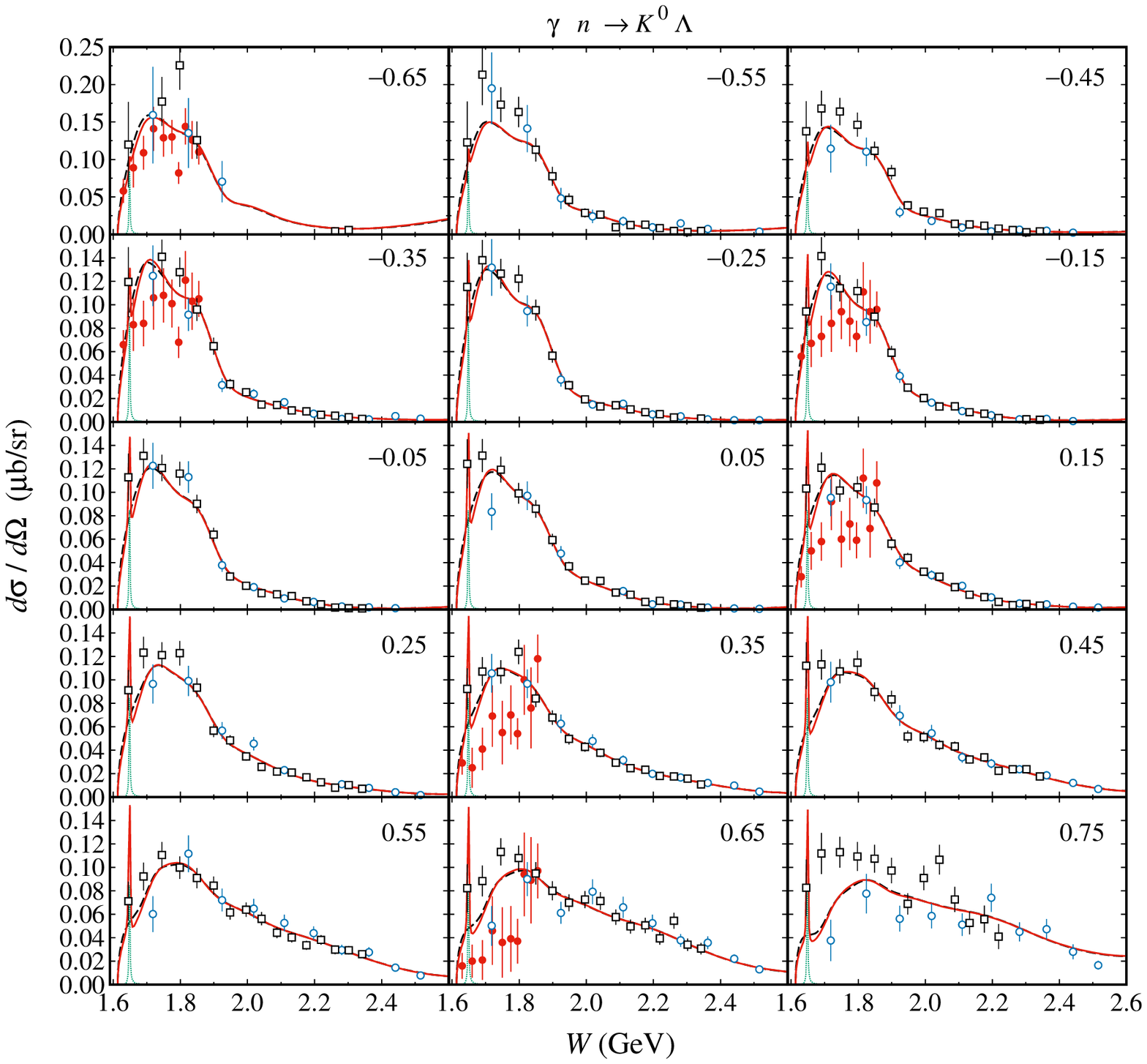}
\caption{Effects of the $P_{11}$ narrow nucleon resonance on the differential cross 
  section of the $\gamma n\to K^0\Lambda$ channel. Solid (dashed) curves are obtained 
  by including (excluding) this resonance in the model M3, respectively.
  The dotted curves exhibit contribution of this resonance alone. Experimental data
  are as in Fig.~\ref{fig:k0lam_th}.}
\label{fig:k0lam_nar_w}
\end{figure*}

Comparison between calculated total cross sections for all models and experimental data
is shown in Fig.~\ref{fig:narrow}. It is obvious that for all models the narrow resonance 
signal is very weak in the  $\gamma p\to K^+\Lambda$ channel [panels (a)$-$(c)]. 
This result corroborates
our previous finding obtained by using the same isospin channel \cite{mart-narrow}. 
On the contrary, the signal is very strong in the $\gamma n\to K^0\Lambda$ channel
[see panels (d)$-$(f)]. 
As expected from the values of $r_{N^*}$ given in Table~\ref{tab:narrow} the calculated 
peak is very strong for model M1 [panel (d)]
and very weak for model M2 [panel (e)]. In fact, in the latter 
the narrow resonance does not create a peak in the total cross section. The effect is 
very small and presumably cannot be resolved by the current data accuracy. 

The use of both CLAS and MAMI data (model M3) reduces the resonance peak in the $\gamma n\to K^0\Lambda$ 
total cross section as shown in panel (f) of Fig.~\ref{fig:narrow}. However, different from
model M1, where the peak wildly overshoots the data, in model M3 the resonance peak seems
to be more natural because the calculated total cross section lies between the CLAS g13 
and MAMI data. Incidentally, there is one CLAS g10 datum in this area which can be 
perfectly reproduced by model M3. Thus, we believe that the model M3 yields a more realistic
effect on the cross section.

Different effects of including the narrow resonance in the $\gamma n\to K^0\Lambda$ channel 
appear in the calculated cross section if we compare panels (d) and (f)  of Fig.~\ref{fig:narrow}.
The difference originates from the different sign of the resonance coupling constants, which are
represented by the ratios $r_{N^*}$ of both models (see Table~\ref{tab:narrow}). In panel (d)
the effect is directly constructive, whereas in panel (f) the effect is slightly destructive, 
which is presumably due to the difference phase.

For a further analysis of the narrow resonance effect on the $\gamma n\to K^0\Lambda$ process 
in Fig.~\ref{fig:k0lam_nar_w} we show energy distribution of the differential cross section for
a number of kaon scattering angles. From this figure we can understand that the largest
effect is obtained in the forward direction. The effect gradually reduces as the kaon
angle increases. Therefore, for the investigation of narrow resonance it is important
to measure the $\gamma n\to K^0\Lambda$ differential cross section in the forward direction
with high statistics and $W$ between threshold and 1.70 GeV. Nevertheless, it is also very
important to solve the problem of data discrepancy before we can draw a solid conclusion.

\section{SUMMARY AND CONCLUSION}
\label{sec:summary}
We have extended the result of our previous investigation on the $\gamma p \to K^+\Lambda$ 
photoproduction channel to analyze the new $\gamma n\to K^0\Lambda$ photoproduction data
obtained from the CLAS and MAMI collaborations. To this end, we have used the effective Lagrangian
method and coupled the two photoproduction processes by utilizing the isospin symmetry and
some information from the Review of Particle Properties of PDG. 
To analyze more than 9400 experimental data points we have considered a number of parameters
in the background and resonance amplitudes as free parameters and adjusted their values
by fitting the calculated observables to experimental data. To constrain the resonance parameters
we used the uncertainties given in the PDG estimates. The presented models can nicely reproduce
the $K^+\Lambda$ data and were used to investigate the effects of the data discrepancy found
in the $K^0\Lambda$ channel. Based on the $K^0\Lambda$ data included in the fitting process, i.e.,
the CLAS, MAMI, or both CLAS and MAMI data sets, three different models M1$-$M3, respectively, 
were proposed in the present work. All models can nicely reproduce the $K^+\Lambda$ data.
In the case of the $K^0\Lambda$ channel the agreement of model prediction and experimental data 
depends on the $K^0\Lambda$ data set included during the fitting process. However, 
we found that the new CLAS beam-target helicity asymmetry $E$ for the $\gamma n\to K^0\Lambda$ 
channel can be better explained by the model that fits the $K^0\Lambda$ CLAS differential 
cross section data. We have also scrutinized  the significance of 
each nucleon resonance involved in the models and found that the 
$N(1720)P_{13}$, $N(1900)P_{13}$, and $N(2060)D_{15}$ resonances are important to 
reduce the $\chi^2$. This result is consistent with the finding of  previous works. 
Finally, we investigated the effect of the $P_{11}$ narrow resonance in both isospin channels 
and found that the effect is more apparent in the $\gamma n\to K^0\Lambda$ process. In this case
the model M3 that fits both CLAS and MAMI data yields the most realistic effect. 
To experimentally investigate the effect of this resonance the present calculation recommends 
a measurement of the $\gamma n\to K^0\Lambda$ differential cross section in the forward 
direction and with the energy range $W$ from threshold up to 1.70 GeV.

\vspace{5mm}
\section{ACKNOWLEDGMENTS}
This work has been supported by 
the 2019 Q1Q2 Research Grant 
of Universitas Indonesia, under contract 
No. NKB-0277/UN2.R3.1/HKP.05.00/2019.

\end{document}